\documentclass[a4paper,11pt]{article}
\pdfoutput=1
\usepackage{jheppub}

\usepackage[utf8]{inputenc}
\usepackage{color}
\usepackage{verbatim}
\usepackage{textcomp}
\usepackage{amsmath}
\usepackage{amssymb}
\usepackage{graphicx}
\usepackage{nicefrac}
\usepackage{natbib}

\usepackage{array}
\usepackage{multirow}

\usepackage{slashed}
\usepackage{diagbox}
\usepackage{cancel}
\usepackage[normalem]{ulem}

\newcommand{\rd}[1]{{\color{red} #1}}

\preprint{FTUV-16-0518, IFIC/15-61\\}

\title{\boldmath Higgs lepton flavour violation:  UV completions
 and connection to neutrino masses}

\author[a,b]{\textbf{Juan Herrero-Garc\'ia,}}
\author[b]{\textbf{Nuria Rius}}
\author[b]{\textbf{and Arcadi Santamaria}}

\affiliation[a]{Department of Theoretical Physics, School of Engineering Sciences, KTH Royal Institute of Technology,
AlbaNova University Center, 106 91 Stockholm, Sweden}
\affiliation[b]{Departamento de F\'isica Te\'orica, Universidad de Valencia and IFIC, Universidad de Valencia-CSIC,
C/ Catedr\'atico Jos\'e Beltr\'an, 2 | E-46980 Paterna, Spain}

\emailAdd{juhg@kth.se}
\emailAdd{nuria.rius@ific.uv.es}
\emailAdd{arcadi.santamaria@uv.es}

\abstract{
We study lepton violating Higgs (HLFV) decays, first from the effective field theory (EFT) point of view, and then analysing the different high-energy realizations of the operators of the EFT, highlighting the most promising models. We argue why two Higgs doublet models can have a $\mathrm{BR}(h\rightarrow \tau \mu)\sim 0.01$, and why this rate is suppressed
in all other realizations including vector-like leptons. We further discuss HLFV in the context of neutrino mass models: in most cases it is generated at one loop giving always $\mathrm{BR}(h\rightarrow \tau \mu) < 10^{-4}$ and typically much less, which is beyond experimental reach. However, both the Zee model and extended left-right symmetric models contain extra ${\rm SU(2)}$ doublets 
coupled to leptons and could in principle account for the observed excess, 
with interesting connections between HLFV and neutrino parameters.}


\makeatother

\begin{document}
\maketitle
\section{Introduction}
In the Standard Model (SM) neutrinos are massless and lepton flavours are exactly conserved at all orders in any process; for instance, the Higgs boson cannot decay in two charged leptons of different flavour. On the other hand, we know from neutrino oscillation experiments that neutrinos are not massless and that lepton flavours are not conserved. However, until now, no lepton flavour violation (LFV)  has been observed in processes involving just charged leptons  (as for instance $\mu\rightarrow e \gamma$ or $\mu\rightarrow 3e$) and only very strong upper limits have been set on their branching ratios. This is intriguing, because if lepton flavour is not conserved there should also be LFV in the charged lepton sector at some level. In fact, many low-scale neutrino mass models predict sizable amplitudes for non-oscillatory LFV processes. With the discovery of the Higgs boson the situation has changed dramatically by opening the possibility of testing LFV in Higgs decays. In fact, CMS 8 TeV data show a $2.4 \sigma$ excess in the channel $h\rightarrow \mu \tau$ \cite{Khachatryan:2015kon}, which is translated into a branching fraction:\footnote{CMS 13 TeV data, however, do not seem to observe any significant excess, with a 95\% CL upper limit of ${\rm BR}(h\rightarrow \mu \tau)<1.20\,\%$~\cite{CMS:2016qvi}.}
\begin{equation}
{\rm BR}(h\rightarrow \mu \tau)=(0.84^{+0.39}_{-0.37})\,\%\,,
\end{equation} 
while ATLAS shows no significant deviation $\mathrm{BR}(h\rightarrow \mu \tau)=(0.53\pm 0.51)\,\%$~\cite{Aad:2016blu}, with only a small excess in one of the signal regions that is not statistically significant. This hint at the percent level serves as a motivation and thus we will focus our discussion on the tau-muon channel, although our analysis can be easily extended to any HLFV decay ($h\rightarrow \ell \ell^\prime$).

Indeed, if confirmed, amazingly, this would be the first signal of LFV apart from that seen in neutrino oscillation experiments and, therefore, the SM, and some of its simplest extensions devised to accommodate neutrino masses (plain Dirac neutrinos or see-saw type I), would have to be extended. In fact, even if a Higgs lepton flavour violating (HLFV) signal is predicted in many models, a $\sim 1\%$ branching fraction is too large for most of them, once one takes into account that no other LFV processes have been observed.  

The goal of this paper is studying, with complete generality, which is the type of new physics that could give sizable contributions to $h\rightarrow\mu\tau$ (which in this paper means: $h\rightarrow \mu^+ \tau^- + h\rightarrow \mu^- \tau^+$) in the light of current (and future) LHC measurements.\footnote{Prospects of observing HLFV at a linear collider were studied in ref.~\cite{Kanemura:2004cn}, at a muon-collider in ref.~\cite{Sher:2000uq} and at an $e\gamma$ collider in ref.~\cite{Yue:2015dia}.}
 This process and the constraints imposed by other LFV processes, such as $\tau\rightarrow \mu \gamma$, have been widely studied in many papers. 
Some authors take a pure phenomenological approach, using an effective Lagrangian to explain the HLFV excess and to estimate the expected contributions to other LFV processes. Then,  they employ the limits on these processes to set upper bounds on HLFV~\cite{Goudelis:2011un, DiazCruz:1999xe, Blankenburg:2012ex,Harnik:2012pb, Celis:2013xja, Banerjee:2016foh, Belusca-Maito:2016axk}. Also the connection of HLFV to CP-violating decays has been studied in refs.~\cite{Kopp:2014rva, Hayreter:2016kyv}.

On the other hand, most of the research has focused on analyzing different models which, a priori, could give large contributions to the HLFV process.
 In the following we briefly review the different options that have been studied in the literature. 

Relevant to our work, as they are tree-level prototype examples, are models with vector-like leptons and 
 two Higgs doublet models 
(2HDM) with generic structure (type III). Vector-like leptons motivated by composite Higgs models were studied in ref.~\cite{Falkowski:2013jya}, where too low rates were obtained, $\mathrm{BR}(h\rightarrow\mu\tau)\lesssim 10^{-5}$ (see also ref.~\cite{Altmannshofer:2015esa}). 
Before the CMS hint, 
$h\rightarrow\mu\tau$ in the type-III 2HDM was studied in detail 
in ref.~\cite{Davidson:2010xv},  imposing the upper bounds from $\tau\rightarrow \mu \gamma$, and the sensitivity of LHC to this channel was studied in ref.~\cite{Davidson:2013psa}. This model has been extensively used in order to explain the excess, see for instance refs.~\cite{Sierra:2014nqa,Dorsner:2015mja,Omura:2015nja}, as it can yield $\mathrm{BR}(h\rightarrow\mu\tau)$ values at the $10\%$ level. 
It has also been extended to account for other anomalies: LHCb anomalies, using a  gauge $L_\mu-L_\tau$ symmetry~\cite{Crivellin:2015mga,Altmannshofer:2016oaq} 
and a horizontal gauge symmetry~\cite{Crivellin:2015lwa}, and the recent diphoton anomaly observed by ATLAS and CMS at $\sim 750$ GeV,  by adding new vector-like fermions  in ref.~\cite{Bizot:2015qqo} (see also ref.~\cite{Han:2016bvl}).

Within particular models, different flavour structures have been considered, such as  
minimal flavor violation~\cite{Dery:2013aba,Dery:2014kxa,He:2015rqa,Botella:2015hoa,Baek:2016pef}, natural flavour conservation~\cite{Dery:2013rta}, Frogatt Nielsen models~\cite{Dery:2013rta,Dery:2014kxa, Huitu:2016pwk}, flavor symmetries~\cite{Heeck:2014qea} and discrete symmetries~\cite{Campos:2014zaa,Hernandez:2015dga,Baek:2016kud}.  
Several of the models proposed to explain the anomaly require severe fine-tuning among different parameters, 
in order to cancel too large contributions to charged lepton flavour violation (CLFV)  processes~\cite{Das:2015zwa,Dorsner:2015mja,Cheung:2015yga,Baek:2015mea}.

Many scenarios in which HLFV appears at one loop have been also studied:
 heavy right-handed neutrinos which  generate radiatively light neutrino masses~\cite{Pilaftsis:1992st}, 
 type-I see-saw~\cite{Arganda:2004bz}, inverse see-saw~\cite{Arganda:2014dta, Arganda:2015naa},
 MSSM~\cite{DiazCruz:1999xe, DiazCruz:2002er,Brignole:2003iv, Parry:2005fp,DiazCruz:2008ry, Crivellin:2010er, Giang:2012vs,Arana-Catania:2013xma, Aloni:2015wvn,Alvarado:2016par,Hammad:2016bng}, lepton flavor dark matter~\cite{Phan:2016ouz}, supersymmetry without R-parity~\cite{Arhrib:2012mg} and in the context of Little Higgs models~\cite{Lami:2016mjf}.
 Generically, $\mathrm{BR}(h\rightarrow\mu\tau)$ in these models is very small, ranging from $10^{-9} - 10^{-4}$.

Finally some papers use both, effective Lagrangians and some specific models, to study the HLFV excess and the constraints from other LFV processes~\cite{Dorsner:2015mja,deLima:2015pqa,Buschmann:2016uzg}, but to our knowledge no complete exhaustive analysis of all the possibilities that could give rise to the HLFV anomaly has been performed. Furthermore, its connection to neutrino masses has not been studied thoroughly, but only in specific models, which yield HLFV at one loop.

The effective Lagrangian approach is very general because it includes many models but it is only valid when the masses of the new particles are above the electroweak scale. However, we are looking for an effect which is not a small correction with respect to the SM, as if confirmed $\mathrm{BR}(h\rightarrow \mu\tau)$ is of the same order of magnitude than $\mathrm{BR}(h\rightarrow \tau\tau)$. Then, the assumption that the new particles are much above  the electroweak scale does not need, necessarily, to be true. Moreover, the calculations of other LFV processes in the effective field theory are just estimates. In fact, as we will see explicitly in the case of theories with vector-like fermions, there can be additional contributions much larger than the ones computed in the EFT. Explicit models, on the other hand, are complete and, therefore, by using them one can compute accurately the relations between the different LFV processes. However, of course, the drawback is that the conclusions can only be applied to the specific model studied.         

Here we will bridge the two approaches which, in fact, are complementary, and we will try to obtain general conclusions on the type of new physics that can explain the HLFV excess without conflicting with the limits on other LFV processes. First we will review the EFT theory approach and the different operators that can lead to HLFV. There are several operators of this type, but since some of them can be rewritten in terms of the others by using the equations of motion (or field redefinitions) usually only the simplest of them is considered. However, as we will see, the implications for new physics of the different operators could be quite distinct, as they can be subject to different phenomenological constraints, and therefore we will discuss all of them.  
Next we will classify all the ways of obtaining these operators from exchange of heavy particles at tree level in renormalizable theories. Then, in the renormalizable theories we will analyze the HLFV excess and the constraints from other LFV processes (particularly $\tau \rightarrow \mu \gamma$) to set the viability of the different models to explain the HLFV excess.

We will also consider the possibility of having large HLFV generated at one loop by exchange of heavy particles. 
Finally, since one of the main motivation for LFV are neutrino masses, we will discuss, in view of our previous analysis, the implications of some of the most popular models of neutrino masses for HLFV, highlighting the most promising ones.

The paper is structured as follows. In section~\ref{sec:EFT} we introduce HLFV from an EFT perspective. In section~\ref{sec:tree} we list all possible tree-level renormalizable realizations of the relevant operators, with at most two new particles, and estimate their contribution to HLFV once the constraints from other LFV processes are taken into account. In section \ref{sec:loop} we discuss some ways of generating the HLFV operators at one loop and show that in general they will give too small contributions. In section~\ref{sec:nus} we study the implications of several of the most popular neutrino mass models for HLFV. Finally we summarize and give our conclusions in section~\ref{sec:conc}.

\section{An EFT approach to HLFV} \label{sec:EFT}

The Lagrangian for the SM leptons reads
\begin{equation}
\label{eq:yukawa}
\mathcal{L}_{\rm leptons} \equiv \mathcal{L}_{\rm kin}+\mathcal{L}_{\rm Y}= i\overline{L}\slashed D L+i\overline{e_{\rm R}} \slashed D e_{\rm R}-\left(\overline{L} Y^{\prime}_{\rm e} e_{\rm R}\Phi+ {\rm H.c.}\right)\,,
\end{equation}
where $L$ is the lepton $\rm SU(2)_{L}$ doublet, $e_{\rm R}$ the fermion singlet, $\Phi$ is the Higgs $\rm SU(2)_{L}$ doublet and $Y^{\prime}_{\rm e}$ a general complex Yukawa matrix in flavour space (flavour indices have been suppressed). The covariant derivative $D_\mu$ takes into account the transformation properties of the fields under the SM gauge group:
\begin{equation}  \label{covariant}
	D_\mu = \partial_\mu - i g \, T_a W^a_\mu - i g^{\, \prime} \, Y \, B_\mu\,,
\end{equation}
for a general field that transforms non-trivially under $\mathrm{SU(2)_\mathrm{L}}$ (with coupling $g$), and has hypercharge $Y$ (with coupling $g^\prime$).  $T_a$ are the generators of weak isospin $\mathrm{SU(2)_\mathrm{L}}$, with $T_a=\frac{1}{2} \, \sigma_a$ for  $\mathrm{SU(2)_\mathrm{L}}$ doublets, where $\sigma_a$ ($a=1,2,3$) are the Pauli matrices. 

In the SM  $Y^{\prime}_{\rm e}$ can always be taken diagonal and, if there are no right-handed neutrinos, lepton flavours are exactly conserved. In the SM with right-handed neutrinos, although $Y^{\prime}_{\rm e}$ can still be chosen diagonal, the Yukawa couplings of neutrinos cannot and, therefore, lepton flavours are not conserved. Thus, despite Higgs couplings to charged leptons are diagonal at tree level, there is a contribution to HLFV at the one-loop level, although the prediction is far from experimental sensitivities  \cite{Arganda:2004bz}, similarly to the results for 
other processes with CLFV as $\mu\rightarrow e \gamma$ or $\mu \rightarrow e e e$. However, in some extensions of the SM, HLFV could be generated at tree level and be relatively large. If all the new particles responsible for HLFV are much heavier than the electroweak scale and decouple, one can describe HLFV, with full generality, by using an effective Lagrangian including gauge invariant operators built with SM fields with dimension higher than four and whose effects are suppressed by powers of the scale of new physics \cite{Coleman:1969sm,Weinberg:1978kz,Weinberg:1980wa} (see also refs.~\cite{Buchmuller:1985jz,Bilenky:1993bt,Georgi:1994qn,Wudka:1994ny,Pich:1998xt,Manohar:1996cq}, and references therein). 
The lowest dimension effective operators that can give rise to HLFV have dimension six and, therefore, the effective low energy Lagrangian can be written as 
\begin{equation}\label{eq:hlfv-lagrangian}
{\cal L}_\mathrm{HLFV}=-\sum_a \frac{C_a}{\Lambda^2}{\cal O}_a + \mathrm{H.c.}\,,
\end{equation}
where family indices in $C_a$ have been suppressed. 

There are also dimension-8 operators that can be constructed by adding the singlet $(\Phi^\dagger\Phi)$ to the previous ones. In this case, however, more new fields and couplings are needed to generate them, suppressing necessarily the contribution to HLFV operators (observables) by $\sim v^2/\Lambda^2$ ($\sim v^4/\Lambda^4$) with respect to dimension 6 ones. Therefore, 
for scales above $\sim 4\pi v\sim 3$~TeV, dimension 6 operators are dominant with respect to the dimension eight ones, and in general the EFT is justified whenever the new physics scale is 
$ \Lambda \gg v$, so in the following we will restrict our discussion to dimension-6 operators.

\begin{figure}
	\centering
        \includegraphics[scale=0.55]{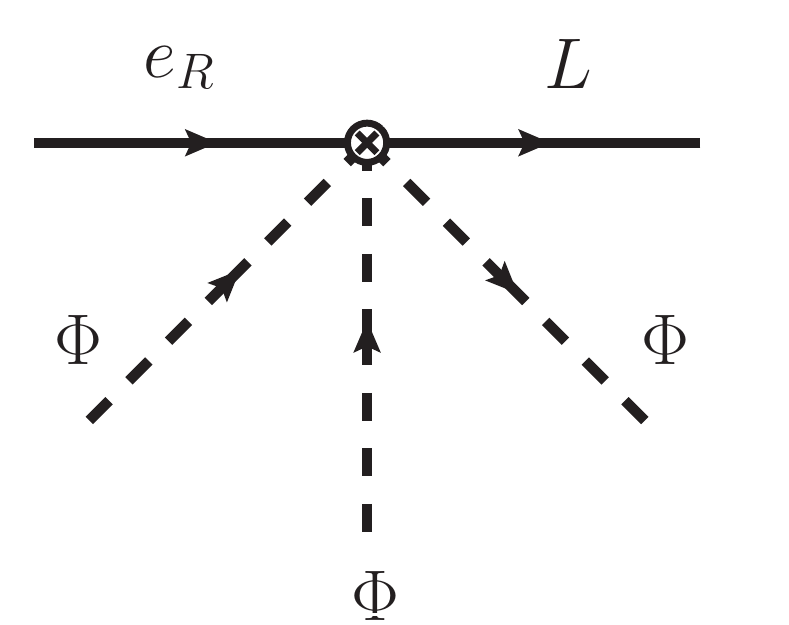}
	\caption{Diagrammatic representation of ${\cal O}_Y$. We use the symbol $\otimes$ to represent effective operator couplings.} \label{fig:effective-yukawa}
\end{figure}
There are many dimension-six operators relevant for HLFV processes, for instance (see figure~\ref{fig:effective-yukawa} for a diagrammatic representation of ${\cal O}_Y$) 
\begin{eqnarray}
{\cal O}_Y &=& \overline{L}e_{\rm R}\Phi(\Phi^\dagger \Phi)\, ,    \label{eq:OH}\\
{\cal O}_\mathrm{1L} &=& (\overline{L}i \slashed D L)\,(\Phi^\dagger \Phi)\, , \,\, 
{\cal O}_\mathrm{1R} = (\overline{e_{\rm R}}i \slashed D e_{\rm R})\,(\Phi^\dagger \Phi)\, ,  \label{eq:O1}\\
{\cal O}_\mathrm{2L} &=& (\overline{L} \tilde{\Phi}) i\,\slashed D  (\tilde{\Phi}^\dagger L)\, , \; \; \;  
{\cal O}_\mathrm{2R} = (\overline{e_{\rm R}} \Phi^\dagger) i\,\slashed D (e_{\rm R}\, \Phi)\, , \;  \, \label{eq:O2}\\
\cdots & \nonumber
\end{eqnarray}
There could also be operators with two derivatives and one Higgs doublet or with three derivates, which could induce HLFV after the use of equations of motion (EOM). However, those operators are more difficult to generate at tree level, they give, in general, more tau-lepton mass suppression factors and, in some cases, they could induce directly $\tau \rightarrow \mu \gamma$. Therefore, they will not be discussed in the following.

The operators in eqs.~(\ref{eq:OH}--\ref{eq:O2}) can be related among them by field redefinitions, integrations by parts or ${\rm SU(2)}$ Fierz identities. For instance, by redefining $L$ as (for simplicity we take the couplings $C_{\rm 1L}$ Hermitian in flavour)  
\begin{equation}
L \rightarrow \Big(1+2\,C_{\rm 1L}\,\frac{\Phi^\dagger \Phi}{\Lambda^2}\Big)^{-1/2} L = \Big(1-C_{\rm 1L}\,\frac{\Phi^\dagger \Phi}{\Lambda^2}\Big) L + \mathcal{O}\Big(\frac{\Phi^\dagger \Phi}{\Lambda^2}\Big)^2\, , 
\end{equation}
one can immediately remove the operator ${\cal O}_\mathrm{1L}$ from the Lagrangian in favour of ${\cal O}_Y$ (which arises when the redefinition is used in the Yukawa Lagrangian, eq.~\eqref{eq:yukawa}) and other operators that do not give HLFV \cite{Harnik:2012pb}
(see ref.~\cite{Corbett:2012ja} for the complete list of EFT operators with the Higgs boson). Notice that this redefinition of the fields is equivalent to having used the EOM to rewrite the operator with derivatives, ${\cal O}_\mathrm{1L}$  in terms of the one  without them, ${\cal O}_Y$.

Therefore, in principle,
one can keep only the effective operator ${\cal O}_Y$ in eq.~\eqref{eq:OH}, as it encodes all the relevant physics for HLFV.
 If one does so, charged leptons obtain their masses from two sources: the standard Yukawa couplings, eq.~\eqref{eq:yukawa}, and the operator ${\cal O}_Y$.
The same two terms also give the couplings to the Higgs boson, $h$, but there is a mismatch between the two coefficients (we use $v=\sqrt{2}\langle \Phi \rangle \approx 246~$GeV)
\begin{equation}
\mathcal{L}_{\rm Y}+\mathcal{L}_{\rm HLFV}=-\overline{e_L}\left(\frac{v}{\sqrt{2}}\left(Y^\prime_e+C_Y\frac{v^2}{2\Lambda^2}\right)
+\frac{h}{\sqrt{2}}\left(Y^\prime_e+3C_Y\frac{v^2}{2\Lambda^2}\right)+\cdots\right)e_R+\mathrm{H.c.}\,,
\end{equation}
so that the diagonalization of the charged lepton mass matrix 
\begin{equation}\label{eq:Mcharged}
M=\frac{v}{\sqrt{2}}\left(Y^\prime_e+C_Y\frac{v^2}{2\Lambda^2}\right)\,,
\end{equation}
$V^\dagger_L M V_R=\mathrm{diag}(m_e,m_\mu,m_\tau)$, does not imply the diagonalization of the Higgs boson Yukawa couplings. In the mass basis the Higgs interactions read 
\begin{equation}\label{eq:yij-lagrangian}
\mathcal{L}_{\rm h}=-\overline{e_{Li}} y_{ij} e_{Rj}h+\mathrm{H.c.}\,,
\end{equation}
with 
\begin{equation}\label{eq:yij}
y_{ij}= \frac{m_i}{v}\delta_{ij}+(V^\dagger_L C_Y V_R)_{ij}\frac{v^2}{\sqrt{2}\Lambda^2}\,,
\end{equation}
where the second term is not diagonal in general and gives rise to HLFV processes.
Notice that the second term in $M$ in eq.~\eqref{eq:Mcharged}, for $\Lambda\gg v$, can be considered as a perturbation to the diagonal $Y_e^\prime$ term, thus  we will have that both $V_{L,R} \approx 1+{\cal O}(v^2/\Lambda^2)$ and then $V^\dagger_L C_Y V_R\approx C_Y +{\cal O}(v^2/\Lambda^2)$. We also define $y_\tau \equiv y_{\tau\tau} \approx m_\tau/v$.

From the interaction in eq. \eqref{eq:yij-lagrangian} one obtains
\begin{equation}\label{eq:BRhtaumu}
\mathrm{BR}(h\rightarrow \tau\mu) = \frac{m_h}{8\pi \Gamma_h} \bar{y}^2\approx 1200\, \bar{y}^2, \qquad 
\end{equation}
where 
\begin{equation}\label{eq:ybardef}
\bar{y}\equiv \sqrt{|y_{\tau\mu}|^2+|y_{\mu\tau}|^2} 
 \simeq \overline{C}_Y \frac{v^2}{\sqrt{2} \Lambda^2} \ , 
\qquad
\overline{C}_Y \equiv \sqrt{|(C_Y)_{\tau\mu}|^2+|(C_Y)_{\mu\tau}|^2} \ .
 \end{equation}
In eq.~\eqref{eq:BRhtaumu} we have assumed that the Higgs decay width is approximately given by the SM value, $\Gamma_h=4.1$ MeV~\cite{Beringer:1900zz}. Then, one can estimate the value of $\bar{y}$ needed to explain the CMS result, 
\begin{equation}\label{eq:ybar}
0.002\,(0.001) <\bar{y}<0.003\,(0.004) \qquad \mathrm{at\ 68\%\ (95\%)\ C.L.}.
\end{equation}
Since $\Gamma_h$ is not completely known (experimentally there are only lower and upper limits) there is some uncertainty in this estimate which, however, only changes slightly the upper limit \cite{Dorsner:2015mja}. Alternatively one can use that the ratio of branching fractions $R_{\mu/\tau}=\mathrm{BR}(h\rightarrow \tau\mu)/\mathrm{BR}(h\rightarrow \tau\tau)=\bar{y}^2/|y_{\tau\tau}|^2$ is independent of the total Higgs decay width. Then, from the experimental limit $0.07<R_{\mu/\tau}<0.21$ (at 68\% C.L.) and assuming that $y_{\tau\tau} \approx m_\tau/v$ one also obtains the ranges in eq.~\eqref{eq:ybar}. 

To obtain such values of $\bar{y}$
with $\overline{C}_Y\lesssim 1$ one will need, from eq.~\eqref{eq:ybardef}, $\Lambda\lesssim 5$~TeV, which is not so far from the electroweak scale. Moreover in many models $\overline{C}_Y \ll 1$, for instance if the tau-lepton Yukawa coupling is the only source of chiral symmetry breaking one expects $\overline{C}_Y \sim m_\tau/v$, or if the operator is generated at one loop one expects a $1/(4\pi)^2$ suppression. In table~\ref{tab:Lambda} we give the natural scale of new physics for the different cases. It is clear that if HFLV is generated at one loop and $m_\tau$ suppressed it will be very difficult to explain the CMS result without conflicting with other experiments,\footnote{Moreover, in this case, in order to have the required values of $\bar{y}$ the scale $\Lambda$ is so low that the EFT approach is not justified.}
and the best chance is for HLFV to be generated at tree level and not suppressed by the tau-lepton mass.
\begin{table}
\begin{center}
\begin{tabular}{c|c}
$\overline{C}_Y$ & $\Lambda\; \mathrm{(TeV)}$ \\ \hline 
1 & 5  \\ \hline 
$m_\tau/v$ & 0.4 \\ \hline 
$1/(4\pi)^2$ & 0.4 \\ \hline
$m_\tau/v(4\pi)^2$ & \rd{0.04}  
\end{tabular}
\end{center}
\caption{\label{tab:Lambda}The scales of new physics necessary to explain the CMS result as a function of the size of the dimensionless coefficient $\overline{C}_Y$ of the effective operator ${\cal O}_Y=\overline{L}e_{\rm R}\Phi(\Phi^\dagger \Phi)$. We highlight in red the case where the EFT approach does not make sense 
for  $\bar{y}$ at the per mille level.}
\end{table}

Before considering the phenomenological constraints on the HLFV couplings $y_{ij}$ it is interesting to estimate their size using naturality arguments. Let us consider the charged lepton mass matrix before diagonalization, $M$, in eq.~\eqref{eq:Mcharged}. For simplicity we will take only two families, $\mu$ and $\tau$. Then, the fact that $M$ is diagonalized by a biunitary transformation implies that $\big|{\det(M)}\big|=\big|M_{\mu\mu} M_{\tau\tau}-M_{\tau\mu} M_{\mu\tau}\big| = m_\mu m_\tau$. If there are no cancellations between the two terms of the determinant, that is, if  $\big|M_{\mu\mu} M_{\tau\tau}\big|\ll \big|M_{\tau\mu} M_{\mu\tau}\big|$ (then $\big|M_{\tau\mu} M_{\mu\tau}\big|\approx m_\mu m_\tau$) or $\big|M_{\mu\mu} M_{\tau\tau}\big|\gg \big|M_{\tau\mu} M_{\mu\tau}\big|$ (then $\big|M_{\tau\mu} M_{\mu\tau}\big|\ll \big|M_{\mu\mu} M_{\tau\tau}\big|\approx m_\mu m_\tau$), we always have $\big|M_{\tau\mu} M_{\mu\tau}\big|\leq  m_\mu m_\tau$. It is also clear that the argument does not work if we allow for cancellations, that is, if $M_{\mu\mu} M_{\tau\tau} \approx M_{\tau\mu} M_{\mu\tau}$. By applying this argument to the explicit form of $M$ in eq.~\eqref{eq:Mcharged} and using that $Y_e^\prime$ can be taken diagonal we can set an upper limit on $\big|(C_Y)_{\tau\mu}(C_Y)_{\mu\tau}\big|$
\begin{equation}
\big|(C_Y)_{\tau\mu}(C_Y)_{\mu\tau}\big|\left(\frac{v^3}{2\sqrt{2}\Lambda^2}\right)^2 < m_\mu m_\tau\,.
\end{equation}
Substituting in eq.~\eqref{eq:yij} we immediately find
\begin{equation}\label{eq:ChengSher}
|y_{\tau\mu} y_{\mu\tau}| \lesssim 4 \frac{m_\mu m_\tau}{v^2} .
\end{equation}
This type of constraints was first proposed in the context of 2HDM and are known under the name of the Cheng-Sher ansatz \cite{Cheng:1987rs} (see also ref.~\cite{Branco:2011iw}). 

This implies that for a symmetric HLFV coupling $y_{\tau\mu}=y_{\mu\tau}$,
\begin{equation} \label{estimate}
\bar{y}=\sqrt{2}|y_{\tau\mu}| \lesssim 2\sqrt{2}\sqrt{\frac{m_\mu m_\tau}{v^2}} = 0.005\,,
\end{equation}
which is compatible with the CMS preferred range, eq.~\eqref{eq:ybar}. However,
the couplings do not need to be symmetric and there could be cancellations among the mass matrix elements. Therefore, we will not impose these constraints.

We have seen that the scale of new physics responsible for HLFV is relatively low (see table~\ref{tab:Lambda}).
This suggests that there could be other LFV processes mediated by the new particles at observable rates.
In fact, the same effective interaction that generates HLFV, eq.~\eqref{eq:yij-lagrangian}, gives $\tau \rightarrow 3 \mu$ at tree level although it is suppressed by the muon mass. There are also one and two-loop contributions to $\tau \rightarrow 3 \mu$, similar to the ones appearing in $\tau\rightarrow \mu\gamma$ but with a virtual photon ``decaying'' to muons, which are dominant because they are not suppressed by the muon mass. An estimate of these diagrams was performed in ref.~\cite{Harnik:2012pb} and compared with the experimental results. The authors conclude that the
 constraints from  radiative decays, $\tau \rightarrow \mu \gamma$, are stronger 
  than the ones obtained from $\tau\rightarrow 3\mu $, mainly because $\mathrm{BR}(\tau\rightarrow 3\mu)$ is suppressed by an additional factor $\alpha$ with respect to $\mathrm{BR}(\tau\rightarrow \mu \gamma)$.
They get  $\bar{y}<0.016$ (at 90\% C.L.), which still allows for HLFV, see eq.~\eqref{eq:ybar}. In fact, if $\bar{y}$ is in the upper region required to explain CMS, $\bar{y}=0.003$, one finds that $\mathrm{BR}(\tau\rightarrow \mu \gamma)\sim 1.5\times 10^{-9}$ to be compared with the present upper limit, $\mathrm{BR}(\tau\rightarrow \mu \gamma)< 4.4\times 10^{-8}$ (90\% C.L.) \cite{Beringer:1900zz}.

\begin{figure}
	\centering
        \includegraphics[scale=0.55]{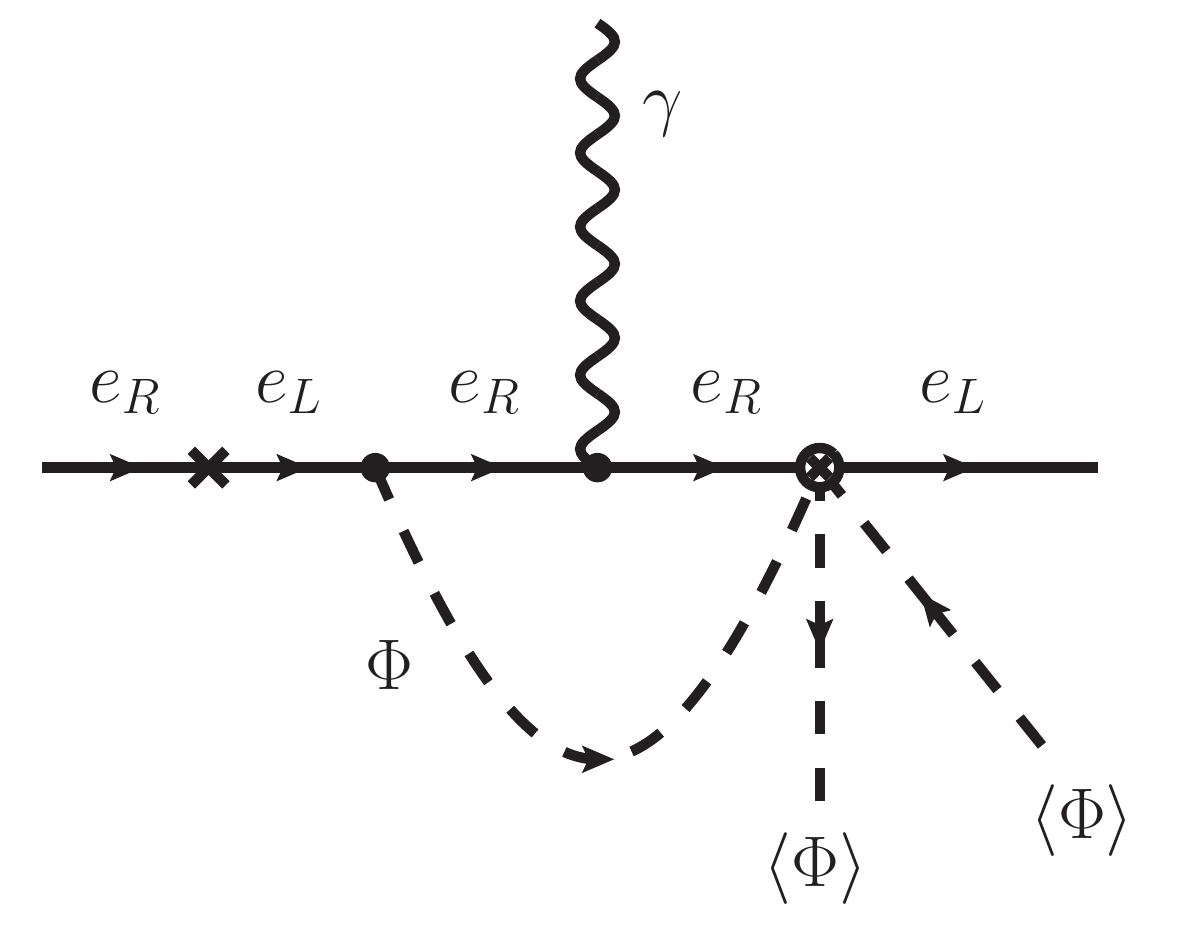}
	\caption{Contribution to $\tau\rightarrow \mu\gamma$ from the effective operator ${\cal O}_Y$. The symbol $\otimes$ represents effective operator couplings, 
the symbol $\bullet$ SM couplings, and $\times$ in external leptonic legs represents a helicity flip produced by $m_\tau$.} \label{fig:effective-gamma}
\end{figure}
It is important to remark that the $\tau \rightarrow \mu \gamma$ (and $\tau\rightarrow 3\mu$) estimates above are calculated in the effective theory, including only the interaction in eq.~\eqref{eq:yij-lagrangian} and only diagrams with SM particles running in loops, see figure \ref{fig:effective-gamma}. In spite that these calculations yield a finite result, in a complete theory there could be other contributions much larger that the ones obtained by computing loops in the effective theory. These contributions to $\tau \rightarrow \mu \gamma$ appear in the EFT as bare effective operators, of dimension six or higher, obtained by matching to the full theory and cannot be computed without knowing the details of the complete theory. It is therefore important to parametrize and, if possible, estimate the form and size of these contributions. 

The simplest operator that gives rise to $\tau\rightarrow\mu\gamma$ is
\begin{equation}\label{eq:ijgamma-operator}
\frac{e C^\gamma_{ij}v}{16\pi^2\Lambda^{2}\sqrt{2}}\bar{e}_{i}\sigma_{\mu\nu}P_{R}e_{j}F^{\mu\nu}+\mathrm{H.c.}\,,
\end{equation}
which appears, after spontaneous symmetry breaking (SSB), from a combination of the gauge invariant operators
$\bar{L}\sigma_{\mu\nu}e_{R}\Phi B^{\mu\nu}$ and $\bar{L}\sigma_{\mu\nu}\vec{\sigma}e_{R}\Phi\vec{W}^{\mu\nu}$. We have  taken into account that this process is always generated at one loop by adding a factor $1/(16\pi^2)$. Notice that, as pointed out in ref.~\cite{Dorsner:2015mja}, the \emph{Yukawa} operator, ${\cal O}_Y$, and these dipole operators have the same transformation properties under flavour and chiral symmetries, so the former will typically give rise to the latter, although it could be suppressed by additional factors. With all this, the $C^\gamma_{ij}$ are expected to be, at most, of order one. Particularizing for the case  $\tau\rightarrow\mu\gamma$ gives 
\begin{equation}\label{eq:tauegamma-operator}
\frac{ev}{16\pi^2\Lambda^{2}\sqrt{2}}\bar{\text{\ensuremath{\mu}}}\sigma_{\mu\nu}\left(C^\gamma_{\mu\tau}P_{R}+C^{\gamma*}_{\tau\mu} P_{L}\right)\tau F^{\mu\nu}\,,
\end{equation}
which leads to the following branching ratio
\begin{eqnarray}
\mathrm{BR}(\tau\rightarrow\mu\gamma) &=&
\mathrm{BR}(\tau\rightarrow\mu\nu\nu) \frac{3\alpha v^2}{8\pi G_{F}^{2}\Lambda^4 m_\tau^2}\left(|C^\gamma_{\mu\tau}|^{2}+|C^\gamma_{\tau\mu}|^{2}\right)\nonumber\\
&\approx& 0.03\left(\frac{\mathrm{TeV}}{\Lambda}\right)^4 \overline{C}_\gamma^2\; ,\;\; \mathrm{with}\;\; {\overline C}_\gamma\equiv  \sqrt{|C^\gamma_{\mu\tau}|^{2}+|C^\gamma_{\tau\mu}|^{2}} \, .\label{eq:Rgamma}
\end{eqnarray}
Since experimentally $\mathrm{BR}(\tau\rightarrow \mu \gamma)< 4.4\times 10^{-8}$, we have 
\begin{equation}\label{eq:Cgamma-bound}
\frac{\overline C_\gamma}{\Lambda^2} \lesssim 10^{-3}\; \mathrm{TeV}^{-2}\; . 
\end{equation}
Then, if $\overline{C}_\gamma \sim 1$ we obtain that $\Lambda > 30$~TeV.
Comparing with table \ref{tab:Lambda} we see that in this case there is no hope. However, $\overline{C}_\gamma$ does not need to be of order one. In fact, in many models the operators generated are not  $\bar{L}\sigma_{\mu\nu}e_{R}\Phi B^{\mu\nu}$ but of the type
 $\bar{L}\sigma_{\mu\nu}\,i\,\slashed{D} L B^{\mu\nu}$, which after the use of the equations of motion can be related to the former but with an additional factor $m_{\tau}/v$. Diagrammatically we say we need a $m_{\tau}$ factor in order to flip chirality. This is always the case for theories with only additional scalars like the 2HDM. Moreover, in some cases, one of the couplings in the loop has to be proportional also to the tau Yukawa coupling, which brings an additional $m_{\tau}/v$ factor. This is the case of the Higgs exchange diagram in the EFT, see figure~\ref{fig:effective-gamma}. Thus, we will consider three classes of models depending on the expected size of the $\overline{C}_\gamma$ coefficients:\footnote{Notice, however, that when the one-loop $\tau\rightarrow \mu \gamma$ amplitude is suppressed by the tau-lepton Yukawa couplings it is possible that two-loop diagrams involving the top quark (the so called Zee-Barr diagrams \cite{Barr:1990vd}) give somehow larger contributions.} 
\begin{equation}\label{eq:Cgammas}
\overline{C}_\gamma \sim 1,\quad \frac{m_\tau}{v},\quad \frac{m^2_\tau}{v^2}\,.  
\end{equation}
With this, one can estimate the expected upper bounds on $\mathrm{BR}(h\rightarrow \tau \mu)$ from $\mathrm{BR}(\tau\rightarrow \mu \gamma)< 4.4\times 10^{-8}$, for models which have different suppressions in $\overline{C}_Y$ and $\overline{C}_\gamma$. In table \ref{tab:CHvsCgamma}, we obtain a lower bound on $\Lambda$ from the $\overline{C}_\gamma$ given in eq.~\eqref{eq:Cgammas} and eq.~\eqref{eq:Cgamma-bound} and use it to estimate an upper bound on $\mathrm{BR}(h\rightarrow \tau \mu)$ for the different possible values of $\overline{C}_Y$, both at tree level and at one loop, and with/without tau-mass suppressions. We conclude that if $\overline{C}_\gamma$ is suppressed by a factor $m_\tau/v$, then $\overline{C}_Y$ should be order one, while if $\overline{C}_\gamma$ is suppressed by a factor $m^2_\tau/v^2$, $\overline{C}_Y$ can be suppressed at most by $m_\tau/v$ or by a one-loop factor, $1/16\pi^2$, but not by both at the same time. Of course, these are just general estimates that perhaps can be avoided in particular fine-tuned models, but at least they give us an idea of what one should expect.

\begin{table}
\begin{center}
\[\text{Upper bounds on } \mathrm{BR}(h\rightarrow \mu\tau)\]
\begin{tabular}{c|c|c|c}
\diagbox[width=7em]{$\overline{C}_Y$}{$\overline{C}_\gamma$} &  $m^2_\tau/v^2$ &  $m_\tau/v$ & 1\\ \hline 
1 & $ 1$  & $ 0.04$ &$ \rd{10^{-6}}$  \\ \hline 
$m_\tau/v$ & $ 0.04$ & $ \rd{10^{-6}}$ & $\rd{10^{-10}}$\\ \hline 
$1/(4\pi)^2$ & $ 0.03$ & $\rd{10^{-6}}$ & $\rd{10^{-10}}$ \\ \hline
$m_\tau/(v(4\pi)^2)$ & $\rd{10^{-6}}$ & $\rd{10^{-10}}$& $\rd{10^{-14}}$ 
\end{tabular}
\end{center}
\caption{The expected upper bounds on $\mathrm{BR}(h\rightarrow \tau \mu)$ from $\mathrm{BR}(\tau\rightarrow\mu\gamma)$ upper limits for different models that give the $\overline{C}_Y$ and $\overline{C}_\gamma$ values in the first column and row respectively. The models outlined are assumed to have no fine-tuned cancellations among different contributions to CLFV, otherwise those bounds could be evaded. The contributions proportional to the tau-lepton mass are generated by \emph{derivative} operators, ${\cal O}_{1L,1R,2L,2R}$, eqs.~\eqref{eq:O1} and \eqref{eq:O2}, while those without it are due to ${\mathcal O}_Y$, see eq.~\eqref{eq:OH}. The $1/(4\pi)^2$ loop factor in $\overline{C}_Y$ explicitly divides the models in tree level and one loop level. In red we highlight the  models unable to explain a $1\%$ $\mathrm{BR}(h\rightarrow \tau \mu)$, unless cancellations occur. \label{tab:CHvsCgamma}}
\end{table}

On the other hand, as we have seen in table \ref{tab:Lambda}, in order to explain the CMS result the scale of new physics cannot be much larger than the electroweak scale and, if it is very close to it, the effective field theory treatment is not appropriate. It is, therefore, very important to scrutinize the complete renormalizable models that could give rise to large HLFV. For this purpose the effective field theory language is still useful since by classifying the different ways of generating the effective operators that give HLFV we are listing in a systematic way the different models. Thus, in the next sections we will classify different ways of ``opening'' the HLFV effective operators with renormalizable interactions.

The different models/topologies are shown in figures~\ref{openopYuk} and \ref{openop}, and explicitly listed in tables~\ref{tab:topologiesH} and~\ref{tab:topologiesE}, respectively. 
Constraints stemming from $\tau\rightarrow\mu\gamma$ are present in all models because all of them contain new charged particles. In general, $\tau\rightarrow\mu\gamma$ will receive several contributions:\footnote{Examples with explicit diagrams will be given when discussing the different topologies in the next section.} 
\begin{description}
\item a) Those which can be calculated in the effective theory (with the Higgs boson and tau-leptons in the loop, see figure~\ref{fig:effective-gamma}), which often are the only contributions considered. These contributions are obtained in the full theory from diagrams which reduce to figure~\ref{fig:effective-gamma} when all propagators with heavy particles shrink to a point
 (see left diagram of figure~\ref{fig:topoC-c} for the case of topology C).

\item b) When integrating the new particles at tree level one could generate additional operators that do not contribute to $h\rightarrow \tau \mu$ but could generate $\tau\rightarrow\mu\gamma$ at one loop (this is for instance the case of operators which induce LFV interactions of the Z bosons in models with vector-like leptons, topologies C, D and E). 

\item c) Direct contributions to $\tau\rightarrow\mu\gamma$ which can only be obtained by matching the EFT to the full theory.
\end{description}

Contributions a) are universal and they are directly linked to $h\rightarrow \tau \mu$, however, they are suppressed by, at least, two $m_\tau$ factors (one factor from the Higgs-lepton coupling and one from the external line helicity flip). Therefore, they will provide the most conservative bound on $h\rightarrow \tau \mu$. 

Contributions b) are model-dependent, because the extra operators generated depend on the particle content.
However, in some cases the parameters appearing in those contributions can be directly linked to the parameters appearing in $h\rightarrow \tau \mu$.
 
Contributions c) are also model-dependent. There are two types of contributions of this class: c1) which can be obtained from the tree level topologies by closing two Higgs doublets and attaching a photon to one of the charged particles in the loop (in the Higgs or in the new scalars/vector-like leptons, depending on the topology). For an example of those contributions in the case of topology C see right diagram of figure~\ref{fig:topoC-c}. These contributions, by construction, depend on the same couplings that appear in $h\rightarrow \tau \mu$ and, therefore, will tightly constrain it. c2) Some of the particles entering in the tree level topologies can independently give contributions to $\tau\rightarrow\mu\gamma$ (for instance in the 2HDM, diagrams with only the non-SM doublet; see also figure~\ref{fig:topoC-c2} for the case of topology C). These can generically be enhanced but contain some parameters which do not appear in $h\rightarrow \tau \mu$. Therefore, they can be set to zero, suppressing $\tau\rightarrow\mu\gamma$ without suppressing $h\rightarrow \tau \mu$.

Upper bounds obtained from a), c1) and in some cases b) will be \emph{robust} because in those cases the $\tau\rightarrow\mu\gamma$ amplitude, $\overline{C}_\gamma$, will be proportional to the $h\rightarrow \tau\mu$ amplitude, $\overline{C}_Y$, and therefore they can not decouple, unless fine-tuned cancellations occur. In the other cases, there will be new couplings involved and, thus, one will be able to set upper bounds on $h\rightarrow \tau\mu$ only under certain extra assumptions. These last limits, which could be avoided in some particular set-ups, will be termed \emph{natural} here. 

\section{Tree level UV completions of the HLFV effective operators} \label{sec:tree}

 As we have discussed in the previous section, despite the fact that all dimension-six operators that give rise to HLFV can be reduced to the ${\cal O}_Y$ operator, the kind of new particles that generate the ${\cal O}_Y$ and ${\cal O}_\mathrm{1L,1R,2L,2R}$ operators could be different and thus so is the phenomenology they induce. For instance, as we will see, operators of type ${\cal O}_{2L,2R}$ in eq.~\eqref{eq:O2} will give rise to HLFV but with a contribution that is always suppressed by the tau-lepton mass. Moreover, they will also generate LFV interactions of the Z boson, which are strongly bounded. In addition, we will also see that, in some cases,  the ${\cal O}_Y$  will always appear together with some of the other operators. Therefore, we will study first the different ways of obtaining the operator  ${\cal O}_Y$ from tree-level exchange of new particles with renormalizable interactions.  Then, we will also discuss the operators with one covariant derivative and two Higgs doublets (type  ${\cal O}_{2L,2R}$). In both cases we will estimate the expected size of the HLFV decay, the relevant constraints from other processes and discuss some interesting models which realize those operators.

\subsection{The \emph{Yukawa} operator} \label{sec:Yukawa}

\begin{figure}
\begin{tabular*}{1\textwidth}{@{\extracolsep{\fill}}>{\centering}m{0.5\columnwidth}>{\centering}m{0.5\columnwidth}}
\centering{}\includegraphics[scale=0.55]{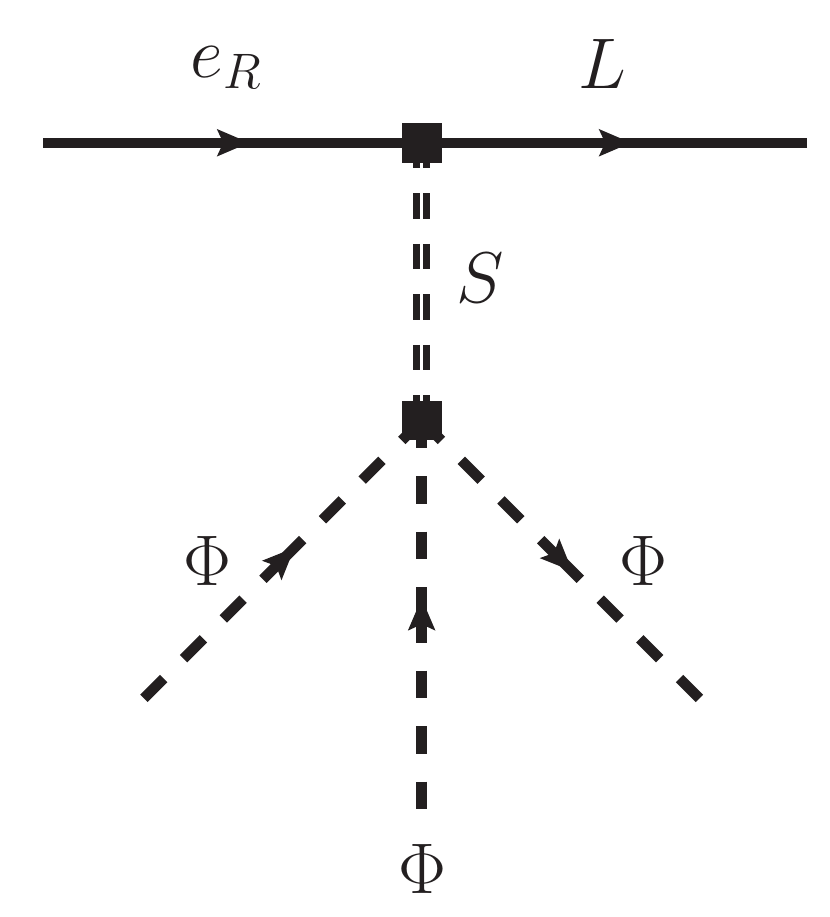}\label{fig:topA} & \centering{}\includegraphics[scale=0.55]{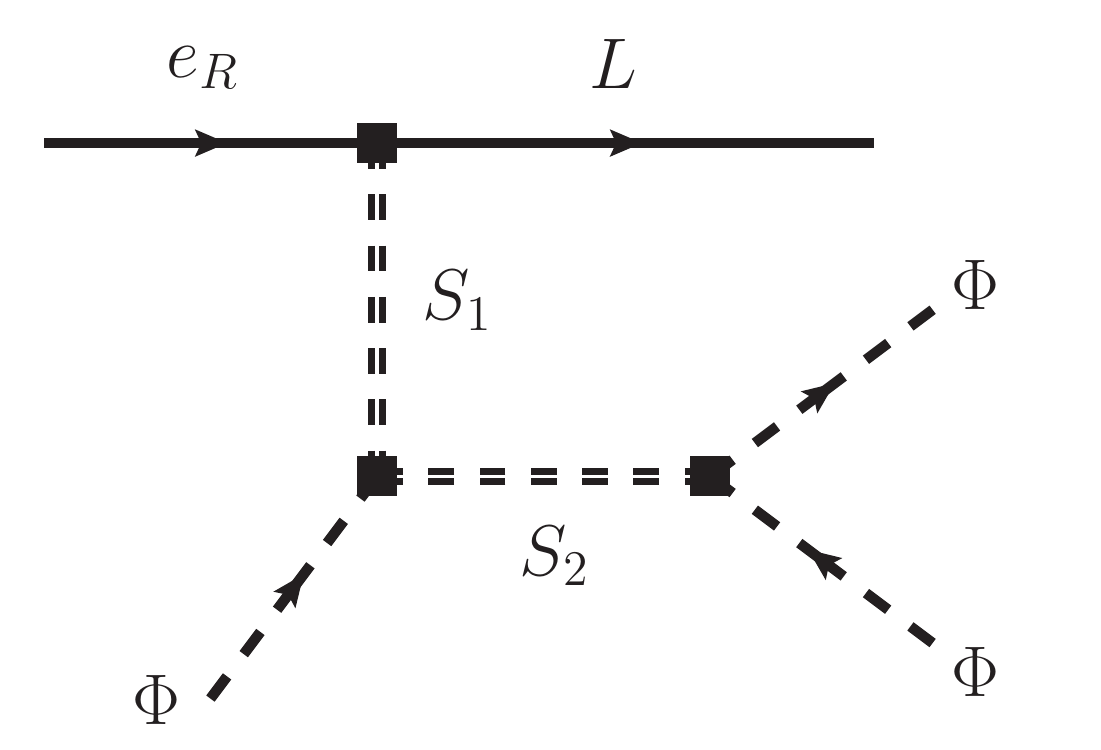}\label{fig:topB}\tabularnewline
(A) & (B)\tabularnewline
\centering{}\includegraphics[scale=0.55]{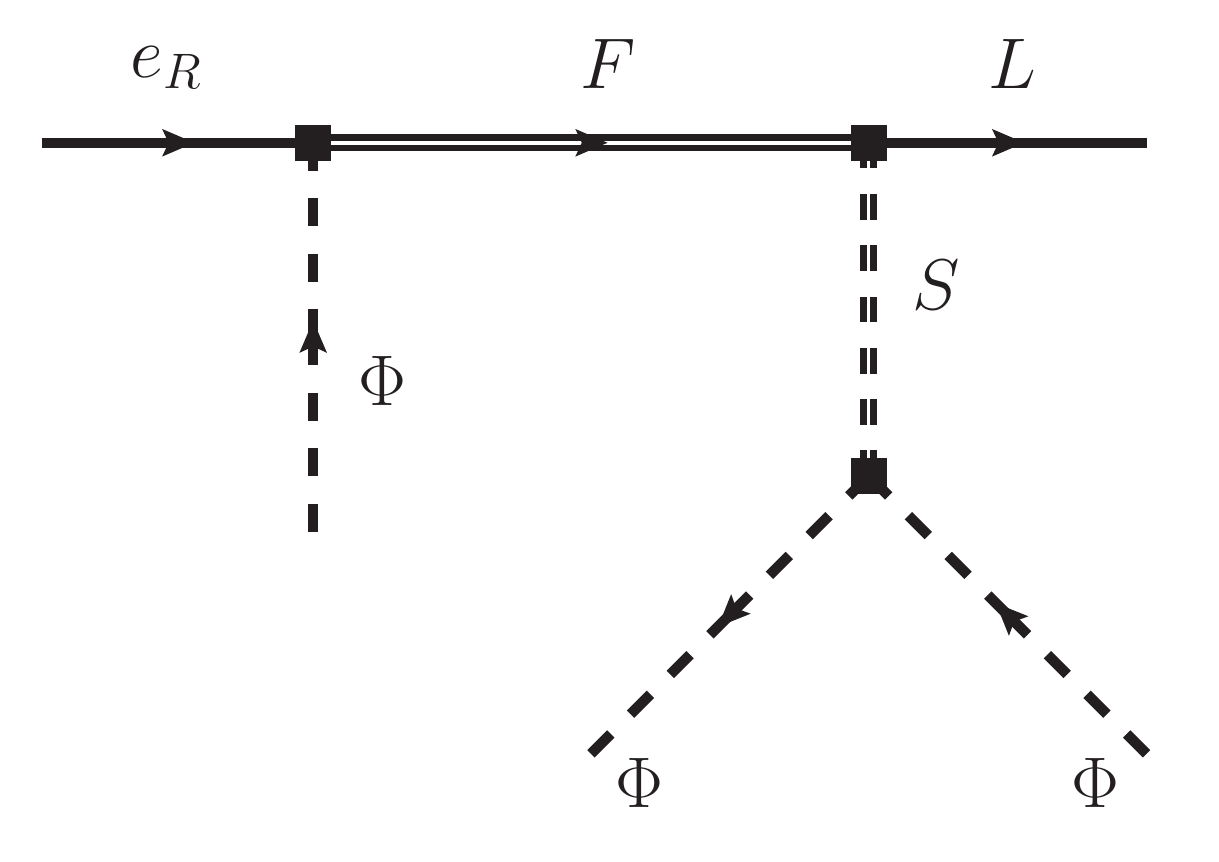} \label{fig:topC}& \centering{}\includegraphics[scale=0.55]{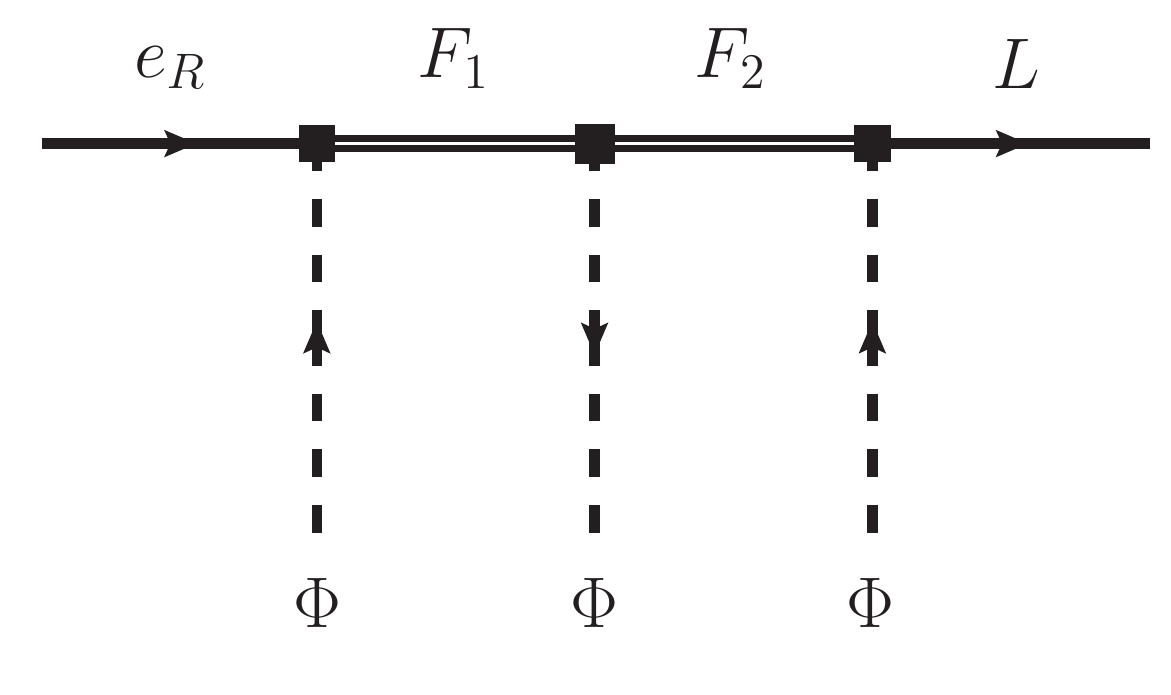}\label{fig:topD}\tabularnewline
(C) & (D)\tabularnewline
\end{tabular*}
\caption{Tree-level topologies that generate the operator ${\cal O}_Y=\overline{L}e_{\rm R}\Phi(\Phi^\dagger \Phi)$, eq.~\eqref{eq:OH}, by exchange of up to two heavy particles (scalars and/or vector-like fermions). Permutations of the three $\Phi$'s give all the possibilities for the quantum numbers of the intermediate particles (see table \ref{tab:topologiesH}).
The symbol $\blacksquare$  represents new physics renormalizable couplings.} \label{openopYuk}
\end{figure}

We give in figure \ref{openopYuk} all the possible tree-level topologies that generate the ${\cal O}_Y$ HLFV operator, eq.~\eqref{eq:OH}, by using renormalizable interactions and including at most two new particles. In table~\ref{tab:topologiesH} we list all the particles that can mediate the interactions in the different topologies,  where S stands for a scalar and F for a vector-like fermion (see refs.~\cite{delAguila:2008pw, deBlas:2014mba}). Fermions must be vector-like because to generate the operator they must be massive before SSB and because they need to flip the chirality.  We use the notation ${\rm (SU(2)_L, Y)}$ and, generically, denote the Yukawa type couplings by $Y$, the scalar quartic couplings by $\lambda$ and the (dimensionful)  trilinear scalar couplings by $\mu$. 

When the scalar $(3,1)_S$ (see-saw type II) or the fermion $(1,0)_F$ (see-saw type I) (or $(3,0)_F$, see-saw type III) are present, (Majorana) neutrino masses could also be generated at tree level.
 Notice however that the representation $(1,0)_F$ (see-saw type I) does not contribute to HLFV at tree level, because it does not contain charged components (shown crossed-out in table~\ref{tab:topologiesH}). 
In principle, the smallness of neutrino masses could impose strong constraints on some of the couplings, however since HLFV preserves total lepton number, it is always possible to avoid the constraints from neutrino masses, 
which violate lepton number and require extra parameters (either couplings or Majorana masses for the 
heavy neutral fermions).  
For instance, the scalar $(3,1)_S$ is present in topologies B and C, and gets a VEV,  but in none of them the coupling of the scalar to two lepton doublets is needed for HLFV. Therefore this coupling could be set to zero avoiding the neutrino mass constraints.
Analogously, although fermions $(3,0)_F$ are present in topologies C, D and E (section \ref{sec:derivative}),
 HLFV relies on the vector-like mass of the fermion. Then, if lepton number is approximately conserved, 
we are  in an inverse see-saw scenario, where neutrino masses can be made small at will, independently of the Yukawa couplings needed to obtain HLFV. In section~\ref{loop} we will discuss the cases of having just the particle content of the see-saws (chiral $(1,0)_F$ or $(3,0)_F$ with Majorana mass terms, or only a $(3,1)_S$ coupled to the lepton doublets), in which HLFV is generated at one loop.

\begin{table} 
\centering
\begin{tabular}{ | c | c |  c |c |}
\hline
Topology & Particles & Representations & Coefficient $\frac{C_Y}{\Lambda^2}$\\ \hline \hline 
$A$ &1\,S &$S=(2,-1/2)$& $\frac{Y \lambda}{m_S^2}$ \\ \hline 

$B$ &2\,S &$(2,-1/2)_S$\,$\oplus$\,$(1,0)_S, (3,0)_S, (3,1)_S$& $\frac{Y \mu_1 \mu_2}{m_{S_1}^2m_{S_2}^2}$\\ \hline 

$C_1$ &1\,F,1\,S &$(2,-1/2)_F$\,$\oplus$\,$(1,0)_S, (3,0)_S$& $\frac{Y_F\,Y^S_F \mu}{m_F m_{S}^2}$\cr
$C_2$ &1\,F,1\,S &$(2,-3/2)_F$\,$\oplus$\,$(3,1)_S$& $\frac{Y_F\,Y^S_F \mu}{m_F m_{S}^2}$ \cr
$C_3$ &1\,F,1\,S &$(1,-1)_F$\,$\oplus$\,$(1,0)_S$, $(3,-1)_F$\,$\oplus$\, $(3,0)_S$& $
\frac{Y_F\, Y^S_F \mu}{m_F m_{S}^2}$\cr
$C_4$ &1\,F,1\,S &$(3,0)_F$\,$\oplus$\,$(3,1)_S$& $\frac{Y_F\, Y^S_F \mu}{m_F m_S^2}$\\ \hline

$D_1$ &2\,F &$(2,-1/2)_F$\,$\oplus$\,$(3,0)_F, \xcancel{(1,0)_F}$& $\frac{Y_{F_1}\, Y_{12}\, Y_{F_2}}{m_{F_1} m_{F_2}}$\cr
$D_2$ &2\,F &$(2,-1/2)_F$\,$\oplus$\,$(1,-1)_F, (3,-1)_F$& $
\frac{Y_{F_1}\, Y_{12}\, Y_{F_1}}{m_{F_1} m_{F_2}}$\cr
$D_3$ &2\,F &$(2,-3/2)_F$\,$\oplus$\,$(1,-1)_F, (3,-1)_F$& $
\frac{Y_{F_1}\, Y_{12}\, Y_{F_2}}{m_{F_1} m_{F_2}}$\\
\hline
\end{tabular}
\caption{Tree-level topologies that generate ${\cal O}_Y=\overline{L}e_{\rm R}\Phi(\Phi^\dagger \Phi)$, eq.~\eqref{eq:OH}, see figure~\ref{openopYuk}. S stands for scalars, F for vector-like fermions which transforms under the given ${\rm (SU(2)_L, U(1)_Y)}$ representation. The scratched representation $(1,0)_F$ does not contribute because it does not contain charged components.\label{tab:topologiesH}} 
\end{table}

Taking all mass scales equal for simplicity, i.e., $m_S\sim m_F\sim \mu \equiv m$ (we know that naturality and charge-breaking constraints impose $\mu\lesssim \mathcal{O}(m)$, so we are considering the most favorable scenario), 
the different topologies obey the following hierarchy in terms of the corresponding Yukawa 
couplings 
\begin{equation} \label{eq:hierarchy}
C_Y \propto \Big( \lambda Y \,:\,Y\,:\, Y^2 \,:\,Y^3\Big)\,.
\end{equation}
In all the topologies the new multiplets contain charged particles, therefore CLFV will constrain the Yukawas $Y$
(besides the perturbativity limit, of $ \mathcal{O}(1)$, which also applies to $\lambda$). 
Moreover, we will see that all the topologies which contain vector-like fermions (C and D) also generate the \emph{derivative} operator, as in table \ref{tab:topologiesE}, and will be subject to further constraints.
Therefore we expect that topologies A and B are the least suppressed. In fact, as already said, topology A (2HDM) is known to generate a sizable HLFV rate.

In the following we are going to discuss in more detail the different topologies.

\subsubsection{Topology A}

The only possibility for the intermediate particle in topology A is a scalar doublet. Then, this topology belongs to the class of two Higgs doublet models, in particular to type-III 2HDM. This is because in the effective theory we assume there is a standard Higgs doublet $\Phi=(2,1/2)$ which gives masses to the fermions and a new one $S$, which generates the HLFV effective operator. Therefore, the two doublets couple to ordinary leptons. 
From the topology A diagram and eqs.~\eqref{eq:hlfv-lagrangian}-\eqref{eq:OH} we find that
\begin{equation}\label{eq:CHtopoA}
\frac{C_Y}{\Lambda^2} \approx \frac{\lambda Y}{m_{S}^2}\,,
\end{equation}
with $\lambda$ the $(S^\dagger\Phi)(\Phi^\dagger\Phi)$ coupling in the potential and $Y$ the Yukawa coupling of the new doublet to the leptons. Then $\bar{y}\sim Y \lambda v^2/m^2_{S}$, and values $\bar{y}\sim 10^{-3}$ can easily be obtained with $m_{S} \lesssim$ few TeV and $\lambda,Y\lesssim 1$.

Now, one has to take into account also phenomenological constraints from
$\tau\rightarrow \mu\gamma$. We can estimate the \emph{robust} upper bound by attaching one of the Higgs doublets to the external tau-lepton line in topology A (see figure~\ref{openopYuk}) and a photon in the (now) internal tau-lepton line (contribution of type a). This will bring a $y_\tau$ factor. Moreover, this coupling will flip the tau-lepton chirality, and to flip back the chirality of the tau-lepton we need an extra $y_\tau$ factor. Finally, the behavior of the integrand at small momenta requires a $m_h^2$ factor in the denominator. Therefore, we find 
\begin{equation}
\overline{C}_\gamma \sim \overline{C}_Y\, \frac{v^2}{m_h^2}\, {y_\tau^2}\,,		
 \end{equation}
which relates directly the $\overline{C}_\gamma$ to the $\overline{C}_Y$ coefficient. It agrees with the one obtained in the effective theory by closing the Higgs in a loop and provides an example of a model in the first column and first row in table~\ref{tab:CHvsCgamma}. The estimate is however too naive: first, because the complete calculation contains some $\log(m_h/m_\tau)$ enhancement factors, and second, because due to the $y_\tau^2$ suppression, there are two-loop contributions (Barr-Zee diagrams \cite{Barr:1990vd}\footnote{It is interesting to see the interplay between CLFV and $h\rightarrow \gamma \gamma$: in the two-loop (Barr-Zee) contributions to $\tau \rightarrow \mu \gamma$ the same couplings of the decay $h\rightarrow \gamma \gamma$ enter.}) which are larger. Adding all these contributions and comparing with the experimental limit on $\tau\rightarrow \mu \gamma$ the EFT contribution gives $\bar{y} < 0.016$ (at 90\% C.L.) \cite{Blankenburg:2012ex,Harnik:2012pb}, which still allows for large $\mathrm{BR}(h\rightarrow \tau\mu)$.

In addition, there are diagrams contributing to $\tau\rightarrow \mu \gamma$ with only the new scalar doublet running in the loops. Denoting its Yukawas with the leptons $Y$, we obtain (example of second column and first row in table \ref{tab:CHvsCgamma}):
\begin{equation}
C^\gamma_{\tau\mu} \sim  Y_{\tau \tau}\, Y_{\tau \mu}\, y_\tau\,. 	
\end{equation}
It is obvious that in this case $h\rightarrow \tau\mu$ is not directly proportional to $\tau\rightarrow\mu\gamma$ because $Y_{\tau\tau}$ does not appear in the former. However, under the natural assumption $Y_{\tau\mu} \lesssim Y_{\tau \tau}$, we can obtain an upper bound on the flavor-violating Yukawa:
\begin{equation}
y_{\tau \mu} \sim Y_{\tau \mu}\, \lambda\, \frac{v^2}{\sqrt{2} m_S^2}\lesssim 0.01\, \lambda\,,		
 \end{equation}	
which yields:
  \begin{equation}
\mathrm{BR}(h\rightarrow \tau\mu)\lesssim 0.2\,\lambda^2 \lesssim 0.2\,,		
 \end{equation}
where in the last step we used the perturbativity upper bound on the quartic coupling, $\lambda\lesssim \mathcal{O}(1)$. Notice that $\mathcal{O}(1)$ values of this coupling would yield problems with perturbativity or stability close to the EW scale. We call these type of bounds \emph{natural} because they could be evaded in some scenarios (for instance if $Y_{\tau\tau}\sim0$).

These estimates agree with the detailed studies of 2HDM \cite{Davidson:2010xv,Sierra:2014nqa}, which include all the scalar contributions (neutral and charged) and also some Barr-Zee two-loop contributions involving the top quark. In the notation of 2HDM, a large $\lambda$ and small enough scale of new physics, i.e., a second Higgs light enough, is seen by having  $\cos(\beta-\alpha)$ large enough (but close enough to the decoupling limit $\sin(\beta-\alpha)=1$ in such a way that the light Higgs is SM-like), where $\tan\beta=v_2/v_1$ and $\alpha$ is the  mixing angle of the two CP-even Higgs scalars, see ref.~\cite{Davidson:2010xv}. After including all known contributions and imposing all phenomenological constraints a $1\%$ branching fraction can be achieved, and therefore one can explain the excess seen by CMS \cite{Sierra:2014nqa}.

\subsubsection{Topology B}
This topology is similar to topology A but with two new scalars, $S_1$ and $S_2$ with trilinear couplings $\mu_1 S_2 S_1^\dagger \Phi$ and $\mu_2 S_2 \Phi^\dagger \Phi$. In fact, the scalar $S_1$ is always a doublet and in the limit of a very large mass of the scalar $S_2$, which can be a singlet or a triplet, topology B reduces to topology A.
From the diagram we obtain that the effective operator coefficient is
\begin{equation}\label{eq:CHtopoB}
\frac{C_Y}{\Lambda^2} \approx Y\,\frac{\mu_1 \mu_2}{m_{S_1}^2 m_{S_2}^2}\, ,
\end{equation}
which, even for $\mu_{1,2}<m_{S_{1,\,2}}$ and $Y<1$ can easily yield $\mathrm{BR}(h\rightarrow \tau\mu)$ at the percent level. The trilinear couplings proportional to $\mu_2$ imply that, after electroweak symmetry breaking, the new scalars $S_2$ get an induced VEV. Thus, when $S_2$ belongs to a triplet there are additional constraints because the $\rho$ parameter receives tree-level contributions from the triplet VEV $v_T$ and, therefore, it is bounded by:\footnote{The VEVs $v_T$ of the scalar triplets $(3,0)_S$ and $(3,1)_S$ contribute as: $\rho_{(3,0)}= 1+ 4 v^2_T/v^2$ and $\rho_{(3,1)}= (v^2+ 2\,v^2_T)/(v^2+ 4 \,v^2_T)$, respectively. Therefore $(3,0)_S$ gives rise to $\rho>1$, while $(3,1)_S$ yields $\rho<1$. Notice that experimentally $\rho=1.00040 \pm0.00024$~\cite{Beringer:1900zz}.} 
\begin{equation}  \label{vevrho}
\frac{v_T}{v}=\frac{\mu_2\, v}{m_{S_2}^2}\lesssim\frac{5\,\rm GeV}{v}\,,
\end{equation}
which implies
\begin{equation}
\mathrm{BR}(h\rightarrow \mu \tau) \sim 1200\, \Big(Y_{\tau\mu}\frac{\mu_{1}v}{m_{S_1}^2}\,\frac{\mu_2 v^2}{v m_{S_2}^2}\,\Big)^2
\lesssim 0.5\, \frac{Y_{\tau\mu}^2 v^2}{m_{S_1}^2}\,,
\end{equation}
where in the last step we used $\mu_1\lesssim m_{S_1}$ from naturalness.

The discussion of possible constraints from $\tau\rightarrow\mu \gamma$ is, in part, similar to topology A. In particular, the contributions obtained by closing the topology and by exchange of the new doublet are essentially the same.
Moreover, the singlet $(1,0)_S$ and triplet $(3,0)_S$ do not couple to leptons and do not give extra contributions to $\tau\rightarrow\mu \gamma$. The triplet $(3,1)_S$ could, in principle couple to leptons, giving  additional contributions to $\tau\rightarrow\mu \gamma$ and generating 
tree-level neutrino masses. However both, neutrino masses and these contributions to CLFV, depend on a Yukawa, see eq.~\eqref{Lst}, which does not enter in HLFV, and thus these constraints are not \emph{robust} and could be evaded. Therefore, we conclude that topologies B can give sizable contributions to HLFV if the new scalars $S_2$ are singlets, as they are not subject to the aforementioned constraints coming from the $\rho$ parameter.

We want to emphasize that the Higgs doublet present in topology B has the same quantum numbers as the Higgs doublet of topology A, and thus, if the contribution from B is generated, so will in general be that of A, which involves less fields and couplings. Which one dominates will depend on the relative size of couplings and masses.

\subsubsection{Topology C}\label{sec:topoC}

Topology C contains one new scalar, $S$,  and one new fermion, $F$. The scalar must have trilinear couplings with the SM Higgs doublet and the fermion must be vector-like to be able to flip the chirality of the SM fermions to generate the operator. 

From the diagram we obtain
\begin{equation}\label{eq:CHtopoC}
\frac{C_Y}{\Lambda^2} \approx Y_F Y_F^S \frac{\mu}{m_F m_S^2}\,,
\end{equation}
where $Y_F$ is the Yukawa coupling of the SM doublet to the SM fermions and the new fermion, while $Y_F^S$ is the Yukawa coupling of the new scalar to the SM fermions and the new fermion. 
Thus
\begin{equation}
\mathrm{BR}(h\rightarrow \mu \tau) \sim 1200\, \Big(\frac{Y_{\tau F} v}{m_F}\, Y^S_{F\mu}\frac{\mu v^2/m_S^2}{v}\Big)^2\,.
\end{equation}
Theories with vector-like fermions will necessarily generate \emph{derivative} operators (topologies E),
 which give additional contributions to $h\rightarrow \tau\mu$, but these contributions are further suppressed by the tau Yukawa coupling. However, the \emph{derivative} operators give rise to other processes (LFV mediated by Z exchange or violations of unitarity in the lepton mixing) which strongly constrain 
the coupling $Y_F$: $(Y_F v/m_F)^2 \lesssim 0.001$, see section~\ref{sec:derivative}. 
Moreover, from the stability of the minimum of the potential we expect $\mu \lesssim m_S$, so that 
\begin{equation}
\mathrm{BR}(h\rightarrow \mu \tau) \lesssim \left(\frac{Y^S_F v}{m_S}\right)^2\,,
\end{equation}
which can be quite large if $m_S$ is not too heavy or $Y_S$ is not too small.
On the other hand, if $S$ is a scalar triplet that gets a VEV, we obtain much stronger bounds from eq.~\eqref{vevrho} (together with $(Y_{\tau F} v/m_F)^2 \lesssim 0.001$)
\begin{equation}
\mathrm{BR}(h\rightarrow \mu \tau) \lesssim 5\times10^{-4}\; (Y^S_F)^2\,,
\end{equation}
which is too small unless we push the Yukawa coupling close to their perturbative limit.

\begin{figure}[h]
	\centering
        \includegraphics[scale=0.5]{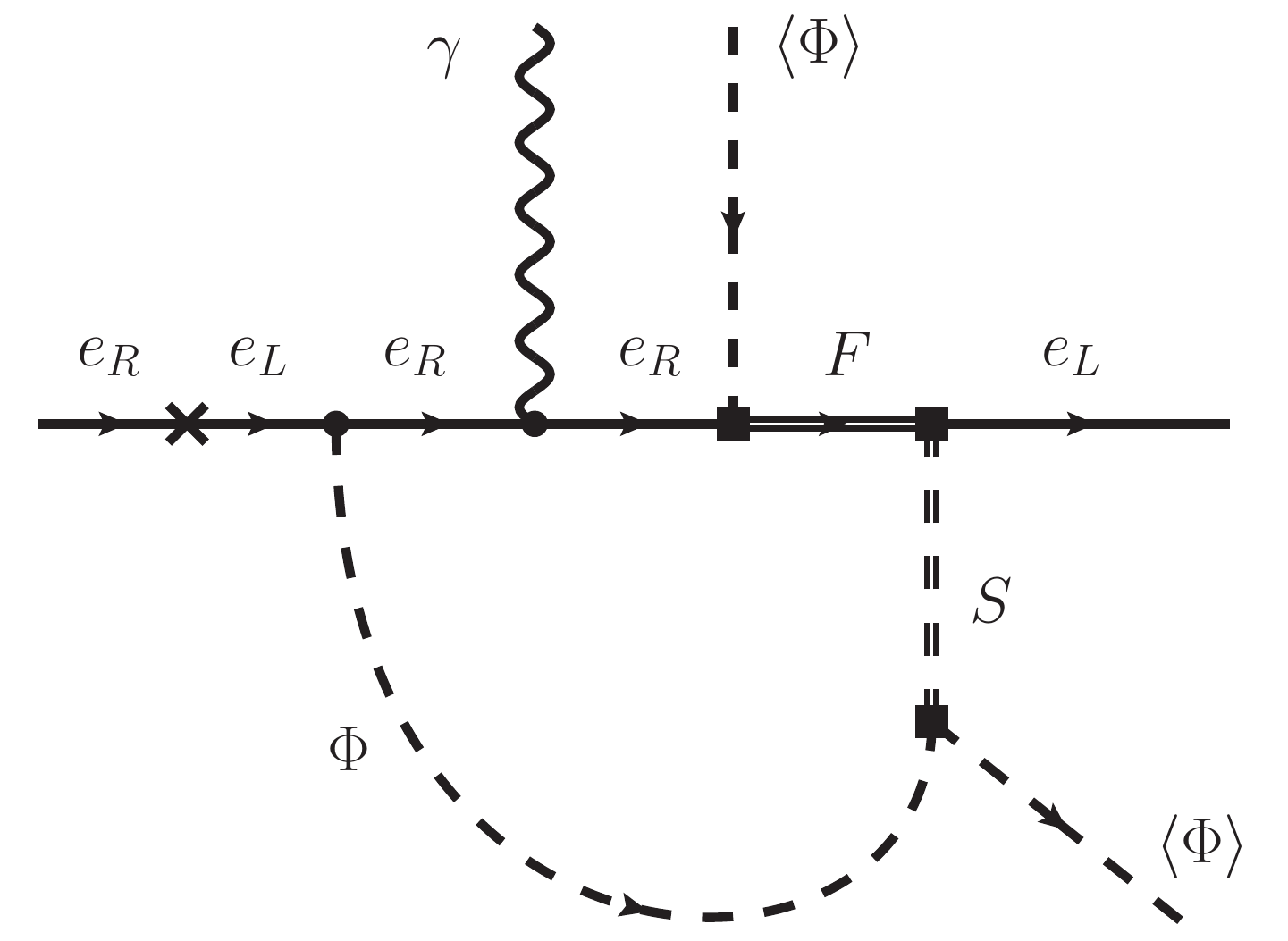} ~~~\includegraphics[scale=0.5]{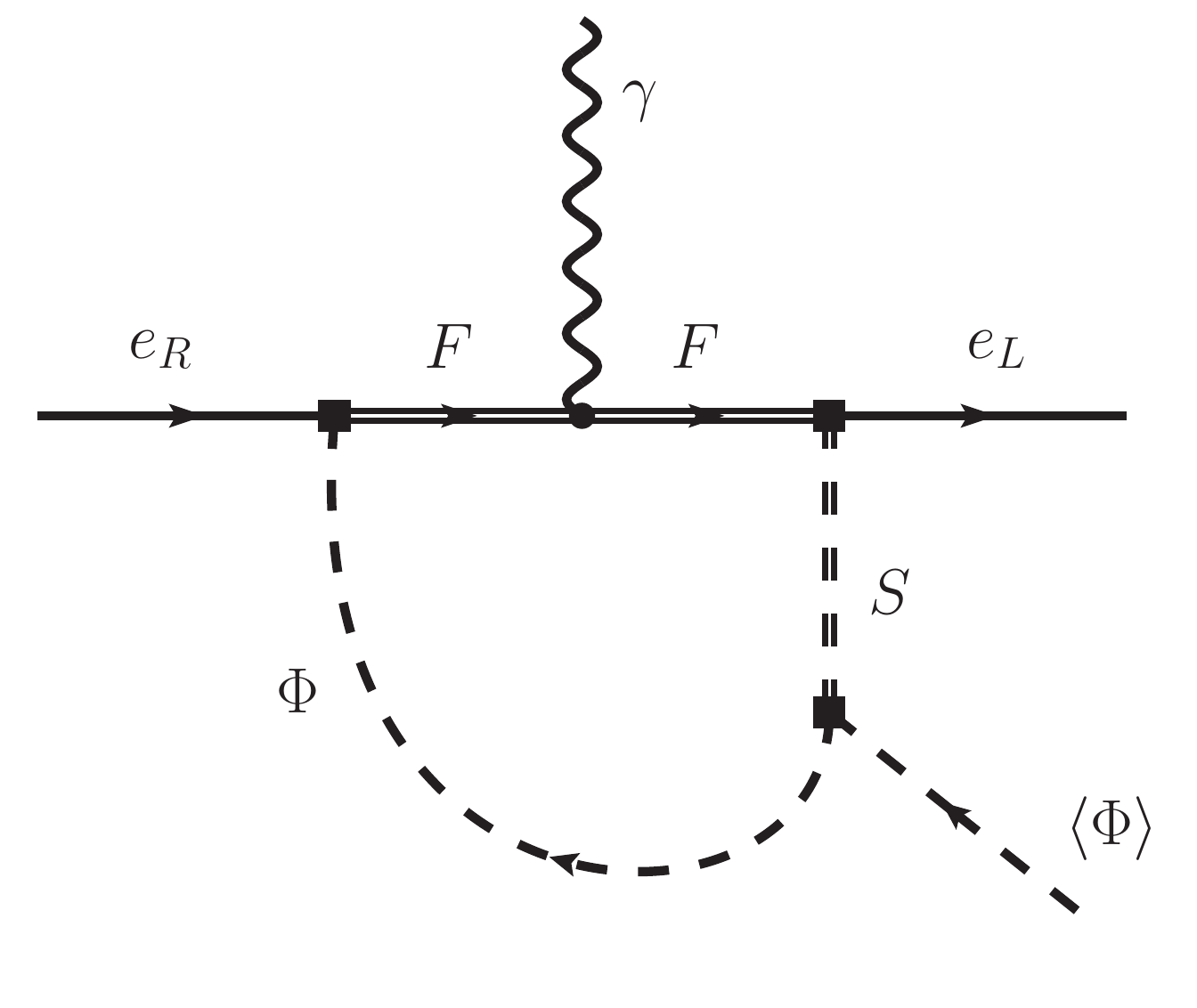}
	\caption{\emph{Robust} contributions to $\tau\rightarrow \mu\gamma$ in topology C.
	Left: It gives rise to the EFT diagram (figure \ref{fig:effective-gamma}) after shrinking the heavy particle propagators (type a contribution). It is suppressed by $m^2_\tau$.
	Right: It cannot be estimated in the EFT (contribution of type c1) and it is not suppressed by $m^2_\tau$.
	The symbol $\bullet$ ($\blacksquare$)  represents SM (new physics) renormalizable couplings,
	 and $\times$ in external leptonic legs stand for a helicity flip produced by $m_\tau$.
	} \label{fig:topoC-c}
\end{figure}
In addition we have to impose the limits coming from $\tau \rightarrow \mu \gamma$. There are many diagrams that give contributions to this process. We will classify them as discussed at the end of section \ref{sec:EFT}. 

For instance, starting from topology C (see figure~\ref{openopYuk}) we can close two of the SM Higgs doublets in a loop and attach a photon to the internal lepton line (see left diagram of figure \ref{fig:topoC-c}). When shrinking the heavy particle propagators to a point this diagram reproduces the ``effective field theory'' result (figure \ref{fig:effective-gamma}). This contribution is what we termed as type a in section \ref{sec:EFT} and, although \emph{robust} in the sense discussed there, it is suppressed by two factors of $m_\tau$. 

If instead we attach the photon to the heavy vector-like fermion, as in the right diagram of figure \ref{fig:topoC-c}, we do not have the $m^2_\tau$ factor because the helicity flip is produced by the heavy vector-like fermion. This contribution, which is also \emph{robust} and is clearly the dominant one, is completely missed in the ``effective field theory'' calculation (type a). In the EFT language it is obtained after matching and can only be estimated after specifying the details of the full theory. This contribution provides and example of what we denoted by type c1 in section~\ref{sec:EFT}. Since it is pure matching, the result is infrared finite and  dominated by the heaviest mass in the loop, so we have:
\begin{equation}
\overline{C}_\gamma\sim \left(\frac{\min(m_F,\, m_S)}{\max(m_F,\,m_S)}\right)^2\,\overline{C}_Y\,.
\end{equation}
To obtain an upper bound on the HLFV rate we  use the lower bounds on the masses of vector-like leptons from direct searches and take $m_F\sim m_S$. Then, using  the limit on $\overline{C}_\gamma$ in eq.~\eqref{eq:Cgamma-bound}, we have that $\bar{y} \sim \overline{C}_Y v^2/(\sqrt{2}\Lambda^2) \lesssim 0.00004$ and from eq.~\eqref{eq:BRhtaumu} we finally get:
\begin{equation}\label{eq:topoC-bound}
\mathrm{BR}(h\rightarrow \tau\mu)\lesssim 2\times 10^{-6}\,.		
 \end{equation}
Notice that this limit can be somewhat relaxed if there is a large hierarchy between the fermion and scalar masses. 

\begin{figure}[h]
	\centering
        \includegraphics[scale=0.55]{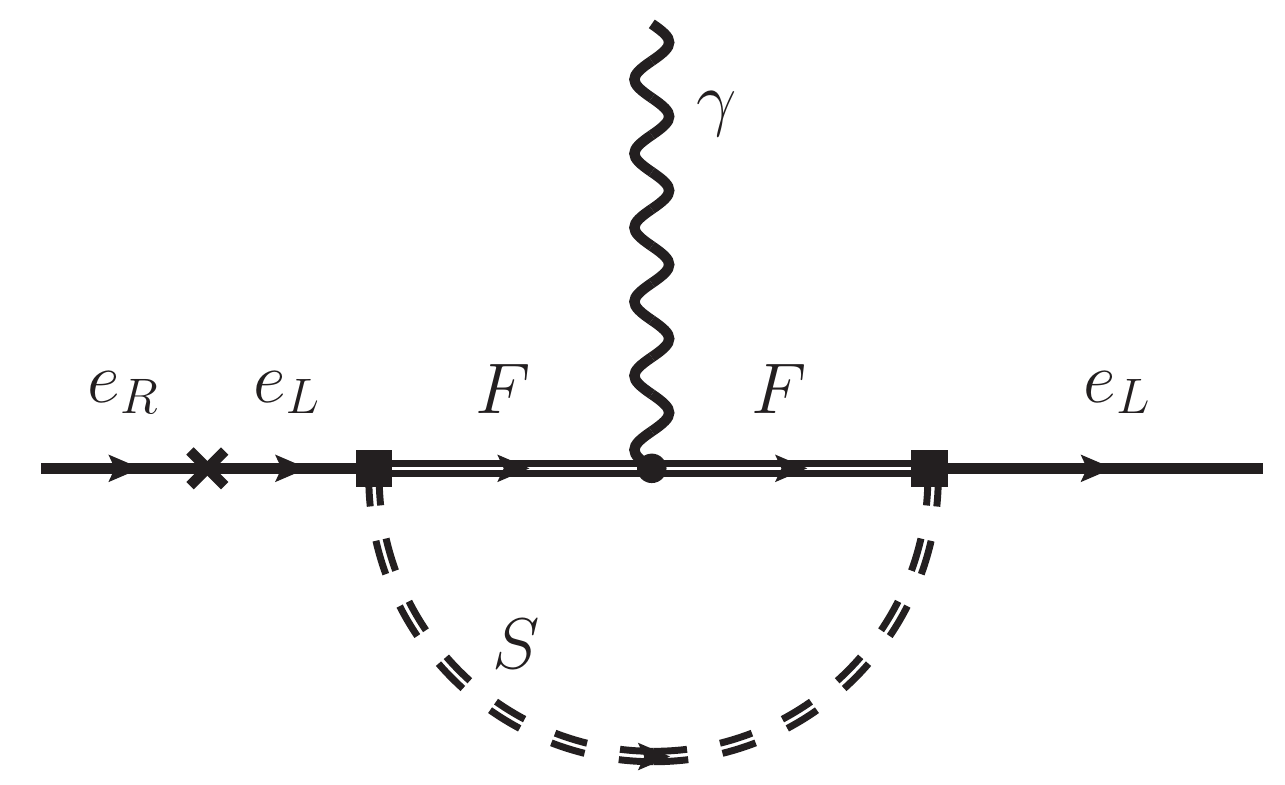}
	\caption{\emph{Natural} contribution to $\tau\rightarrow \mu\gamma$, in topology C, which cannot be estimated in the EFT and contains a different combination of couplings than that appearing in HLFV (contribution type c2). It is suppressed by $m_\tau$.} \label{fig:topoC-c2}
\end{figure}

There are also contributions to $\tau\rightarrow \mu \gamma$ from diagrams in which only the pair new scalar/new fermion is exchanged, see for instance diagram in figure~\ref{fig:topoC-c2} which provides a contribution of the type c2, as defined in section~\ref{sec:EFT}. These necessarily give left-left or right-right amplitudes in the external leptons\footnote{In fact, they generate a derivative CLFV operator like the one discussed after eq.~\eqref{eq:Cgamma-bound}.} and, therefore, require a chirality flip and a factor $m_\tau$.  Typically we  obtain $C^\gamma_{\tau\mu} \propto Y^S_{\tau F}\, Y^S_{\mu F}\, y_\tau$. We can see that, in this case 
$\overline{C}_Y \propto  \mu Y_{\tau F} Y^S_{\mu F}$  is not exactly related to $\overline{C}_\gamma$, which does not contain the factor $\mu$ and contains other Yukawa couplings. One could, for instance, take $Y^S_{\tau F}=0$,  making this contribution to $\mathrm{BR}(\tau\rightarrow \mu \gamma)$ zero while keeping $\mathrm{BR}(h\rightarrow \tau\mu)$ different from zero. Therefore the limits obtained from these diagrams are not \emph{robust}. Moreover, the fact that the amplitudes from these diagrams are proportional to $m_\tau/v$ makes the limits obtained from them a factor $10^4$ weaker which implies that these contributions are irrelevant. 

Finally, there are additional contributions to $\overline{C}_\gamma$ from the derivative operators 
generated by the vector-like fermions, 
but as we will discuss in section~\ref{sec:derivative} these are also $m_\tau$ suppressed.

\subsubsection{Topology D}
These topologies contain two new vector-like fermions $F_1$ and $F_2$ and no new scalars. From the diagram we immediately obtain
\begin{equation}\label{eq:CHtopoD}
\frac{C_Y}{\Lambda^2} \approx  \frac{Y_{F_1} Y_{12} Y_{F_2} }{m_{F_1} m_{F_2}}\,,
\end{equation}
where $Y_{{F_1},\,{F_2}}$ are the Yukawa couplings of the fermions $F_{1,\,2}$ to SM leptons and $Y_{12}$ is the Yukawa coupling among them. Then we have 
\begin{equation}
\mathrm{BR}(h\rightarrow \mu \tau) \sim 1200\, \left(\frac{Y_{\tau F_1} v}{m_{F_1}}\, Y_{12}\,\frac{Y_{F_2\mu}\,v}{m_{F_2}} \right)^2 \lesssim 10^{-3} \,Y_{12}^2 \,,
\end{equation}
where we have used the limits $(Y_{\tau F_1} v/m_{F_{1}})^2,\,(Y_{\mu F_2} v/m_{F_{2}})^2 \lesssim 0.001$ coming from the extra $Z$ interactions generated by the new vector-like fermions (see section \ref{sec:derivative}). Moreover $Y_{12}$ is constrained by present data on  $h\rightarrow \gamma \gamma$ and perturbativity to be $Y_{12} \lesssim {\cal O}(1)$. Therefore using only these arguments it seems difficult to obtain $\mathrm{BR}(h\rightarrow \tau \mu)$ at the percent level.
Regarding the constraints from $\tau\rightarrow \mu \gamma$, we can close the two external SM doublets in a loop and attach a photon to the internal charged particles (contribution c1). We obtain $\overline{C}_\gamma \sim \overline{C}_Y$, as in the case of topology C, and the reason is the same: vector-like fermions provide the required chirality flip. As a consequence, we get  the same bound as in eq.~\eqref{eq:topoC-bound} (example of third column first row in table \ref{tab:CHvsCgamma}),  
\begin{equation}\label{eq:topoD-bound}
\mathrm{BR}(h\rightarrow \tau\mu)\lesssim 2\times 10^{-6}\, ,		
\end{equation}
which is \emph{robust} and renders this topology not useful to obtain an enhancement in $h\rightarrow \tau\mu$. 

As in the case of topology C there are additional contributions both to $\overline{C}_Y$ and $\overline{C}_\gamma$, coming from diagrams with only one vector-like fermion, which are $m_\tau$ suppressed and will be discussed in section \ref{sec:derivative}.

\subsection{The \emph{derivative} operators (topologies E)}\label{sec:derivative}

\begin{figure}
	\centering
        \includegraphics[scale=0.55]{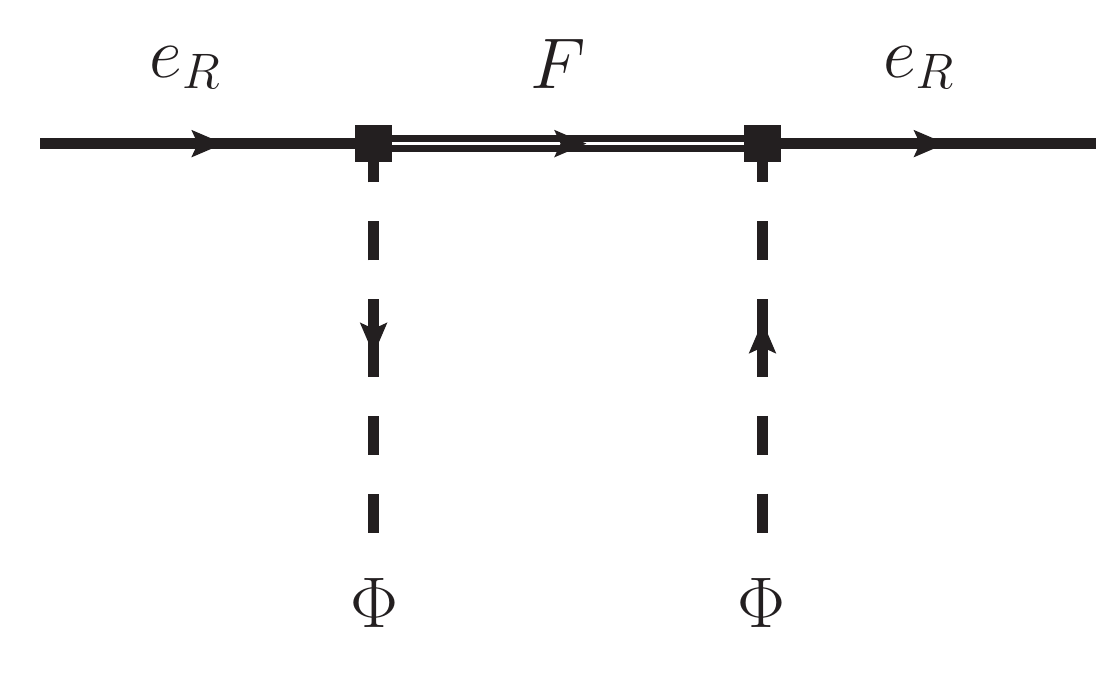}  \qquad
         \includegraphics[scale=0.55]{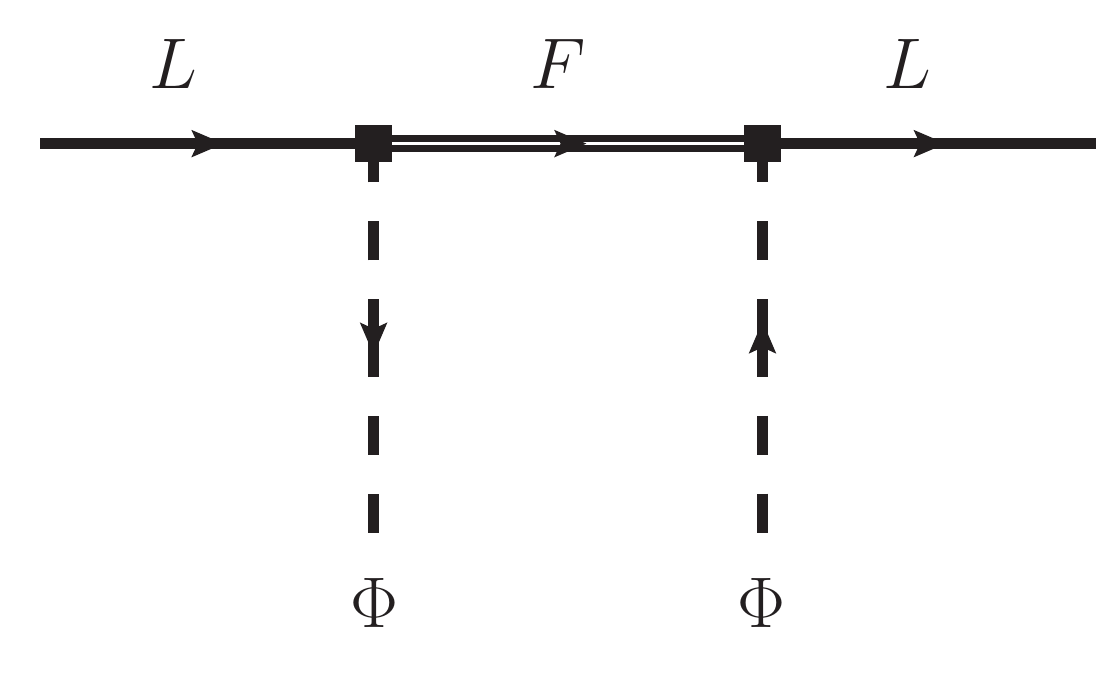} \\
              \vspace{1cm}
     	\caption{Tree-level topologies giving rise to type ${\cal O}_\mathrm{1L,1R,2L,2R}$ operators, see eqs.~\eqref{eq:O1} and \eqref{eq:O2}. Those involving the lepton singlets $e_{\rm R}$ give topologies $E_{1, 2}$ (left diagram) and those involving the lepton doublets $L$ generate topologies $E_{3,4}$ (right diagram). The $\Phi$ and the $\overline{\Phi}$ can be interchanged among the different legs.} \label{openop}
\end{figure}

We show in figure~\ref{openop} all the possible tree-level topologies giving rise to type ${\cal O}_\mathrm{1L,1R,2L,2R}$ operators, see eqs.~\eqref{eq:O1} and \eqref{eq:O2}, by including only one new particle, which must always be a vector-like lepton. In table \ref{tab:topologiesE} we list all the possible high-energy realizations. 

\begin{table} 
\centering 
\begin{tabular}{ | c |c | c |  c |c |c |c |}
\hline
Operator&Topology & Particles & Z $\nu_\alpha \nu_\beta$ & Z $e_\alpha e_\beta$  & W $e\nu$  & H $e_\alpha e_\beta$\\ \hline \hline 
$ (\overline{e_{\rm R}} \Phi^\dagger) i\slashed{D} (e_{\rm R}\, \Phi)$&$E_1$ &$(2,-1/2)_F$&&-1&& 1\\ \hline
$ (\overline{e_{\rm R}} \Phi^T) i\slashed{D} (e_{\rm R}\,\Phi^*)$&$E_2$  &$(2,-3/2)_F$&&+1&& 1\\ \hline\hline

$ (\overline{L} \tilde{\Phi}) i\slashed{D} (\tilde{\Phi}^\dagger L)  $&$E_{3a}$ &$(1,0)_F$ &-1&&-1&   \\ \hline

$(\overline{L} \vec{\sigma}\tilde{\Phi}) i\slashed{D}  (\tilde{\Phi}^\dagger\vec{\sigma}L)   $&$E_{3b}$ &$(3,0)_F$&-1&-2&+1& 2 \\ \hline\hline

$  (\overline{L} \Phi ) i\slashed{D} (\Phi^\dagger\,L) $&$E_{4a}$  &$(1,-1)_F$&&+1&-1& 1\\ \hline

$ (\overline{L} \vec{\sigma}\Phi) i\slashed{D} (\Phi^\dagger\,\vec{\sigma}L) $&$E_{4b}$ &$(3,-1)_F$&+2&+1&+1 & 2\\ \hline
\end{tabular}
\caption{Tree-level topologies of type ${\cal O}_\mathrm{2}$ operators, see eqs.~\eqref{eq:O1} and \eqref{eq:O2}, with the representation under ${\rm (SU(2)_L, U(1)_Y)}$, see figure~\ref{openop}. Higgs couplings are in units of 
$y_{\tau\mu}$,  $Z$ couplings are in units of $\kappa_{\tau\mu} e/(2c_{W}s_{W})$ while $W$ ones are in units of $\kappa_{\tau\mu} e/(2\sqrt{2}s_{W})$ with  $\kappa_{\tau\mu} \sim Y_{\tau F} Y_{\mu F} v^2/(2\,m_F^2)$ and $y_{\tau\mu} \sim  y_\tau \kappa_{\tau\mu}$. The chirality of the charged lepton couplings to the Z are understood from the effective operator. Notice that the couplings of the W and Z to the leptons are related by $g_{W} = g_{ZL\nu} -  g_{ZLe}$.} \label{tab:topologiesE}
\end{table}

After expanding the operators in table \ref{tab:topologiesE} and using the SM equations of motion, they can be written as the operator ${\cal O}_Y$, generating HLFV  suppressed by an additional factor $y_\tau$. 
Therefore, typically they give
\begin{equation}\label{eq:CHderivative}
\frac{C_Y}{\Lambda^2} \sim  y_\tau \frac{Y_{\tau F} Y_{\mu F}}{m_F^2}\; ,
\end{equation} 
and thus $y_{\tau\mu} \sim  y_\tau \kappa_{\tau\mu}$, with  $\kappa_{\tau\mu} \sim Y_{\tau F} Y_{\mu F} v^2/m_F^2$. 
But the same operator will also produce LFV Z-couplings, non-universal tree-level Z-decays and violations of unitarity in the leptonic mixing matrix. We show in table \ref{tab:topologiesE} the new non-diagonal couplings of the different operators (see appendix \ref{sec:nondiag} for the derivation). Higgs couplings are in units of 
$y_{\tau\mu}$, $Z$ couplings are in units of $\kappa_{\tau\mu} e/(2c_{W}s_{W})$ (with $s_W\,(c_{W})$ being the sine (cosine) of weak mixing angle) and $W$ ones are in units of $\kappa_{\tau\mu} e/(2\sqrt{2}s_{W})$. The charged-current interactions are already non-diagonal at the renormalizable level via the PMNS lepton mixing matrix, and therefore this is just a correction to it, i.e., $V\rightarrow V(1+\mathcal{O}(1)\,\kappa\, e/(2\sqrt{2}s_{W})$. The most important constraints on $\kappa_{\tau\mu}$ come therefore from Z non-diagonal couplings.

As we see from table \ref{tab:topologiesE}, in all cases that give HLFV there are non-diagonal Z couplings, which make HLFV negligible if it is generated only by \emph{derivative} operators. For instance, from $\tau \rightarrow 3 \mu$ mediated by the Z at tree level, we find:
\begin{equation} \label{tauuu}
\mathrm{BR}(\tau\rightarrow 3\mu) \sim |\kappa_{\tau\mu}|^2 \mathrm{BR}(\tau\rightarrow \mu \nu \nu) = 0.17\,|\kappa_{\tau\mu}|^2\,.
\end{equation}
Using the experimental limit $\mathrm{BR}(\tau \rightarrow 3\mu)_{\rm exp}<2.1\times10^{-8}$, we get:\footnote{Similar CLFV and non-unitarity constraints are obtained for the $(1,0)_F$ (see-saw type I)~\cite{Abada:2007ux} and $(3,0)_F$ (see-saw type III)~\cite{Abada:2008ea}.}
\begin{equation}  \label{derZ}
|\kappa_{\tau\mu}| \lesssim\mathcal{O}(10^{-3})\,. 
\end{equation}
Since we have seen that  $y_{\tau\mu} \sim  y_\tau \kappa_{\tau\mu}$, we immediately obtain a bound on $|y_{\tau\mu}| \lesssim 10^{-5}$ and therefore
\begin{equation}\label{eq:derivative-bound}
\mathrm{BR}(h\rightarrow \mu \tau) \sim 1200\, |y_{\tau\mu}|^2 \lesssim 10^{-7}\, .
\end{equation}
Moreover, the topologies with vector-like leptons will also give contributions to radiative decays like $\tau \rightarrow \mu \gamma$, which could yield stronger limits. In particular, there are one-loop contributions of type a, b and c1, which, for only one vector-like multiplet, are all suppressed by $m_\tau$. Since, in this case, $C_Y$ is also proportional to $y_\tau$, it provides an example of the second column, second row in table \ref{tab:CHvsCgamma}, and there is no hope to have large $\rm {BR}(h\rightarrow \tau \mu)$. 

Anyway,  the tree-level constraints from LFV in Z-couplings discussed above are strong enough
to prevent sizable contributions to HLFV coming only from topologies of type $E$.
Notice that although such contributions generated by just one vector-like fermion
are also present in topologies $C$ and $D$, in these models 
there are additional HLFV terms which are not suppressed by $y_\tau$ and, thus, the limit from eq.~\eqref{eq:derivative-bound} does not apply.

In appendix~\ref{sec:HiggsVL} we give further details of models with just vector-like leptons giving rise to \emph{derivative} operators, as well as to topology D of the \emph{Yukawa} operator. We show there how, in complete generality, for all vector-like models, Higgs interactions in the mass basis can be unambiguously expressed in terms of Z interactions and are therefore suppressed.

\subsubsection*{Universality of Z decays} 

The leptonic Z-decay branching ratios are universal to a good degree
of precision. In fact from ref.~\cite{Agashe:2014kda} we have 
\[
BR^{\rm exp}_{e}=(3.363\pm0.004)\%\:,\quad BR^{\rm exp}_{\mu}=(3.366\pm0.007)\%\:,\quad BR^{\rm exp}_{\tau}=(3.370\pm0.008)\%\:.\quad
\]
Therefore, we can use them to set limits on violations of $Z$-couplings
universality. If $\delta g_{\ell L,R}^{NP}$ encode all non-universalities
from new physics we have
\begin{equation}
\mu_{\ell}\equiv\frac{BR_{\ell}}{BR_{\ell}^{\rm SM}}=\frac{(g_{L}^{SM}+\delta g_{\ell L}^{NP})^{2}+(g_{R}^{SM}+\delta g_{\ell R}^{NP})^{2}}{(g_{L}^{SM})^{2}+(g_{R}^{SM})^{2}}\approx1+2\frac{g_{L}^{SM}\delta g_{\ell L}^{NP}+g_{R}^{SM}\delta g_{\ell R}^{NP}}{(g_{L}^{SM})^{2}+(g_{R}^{SM})^{2}}\,.\label{eq:mulepton}
\end{equation}
Notice that in the case of the tau-lepton there are small effects
($-0.23\%$) due to the tau-lepton mass, which cancel in this ratio
(but are present in the experimental value $BR^{\rm exp}_{\tau}$). For the SM values we
take $BR^{\rm SM}_{e}=BR^{\rm SM}_{\mu}=(3.366\pm0.003)\%$ and $BR^{\rm SM}_{\tau}=(3.358\pm0.003)\%$.
Then we have 
\begin{equation}
\mu_{e}=0.9991\pm0.0015\:,\quad\mu_{\mu}=1.000\pm0.0023\:,\quad\mu_{\tau}=1.0036\pm0.0025\,.\label{eq:muleptonvalues}
\end{equation}
Thus, being very conservative, we can take for all leptons the $2\sigma$
upper limit obtained for the tau-lepton
\begin{equation}
\left|\frac{2g_{L}^{SM}\delta g_{L}^{NP}}{(g_{L}^{SM})^{2}+(g_{R}^{SM})^{2}}+\frac{2g_{R}^{SM}\delta g_{R}^{NP}}{(g_{L}^{SM})^{2}+(g_{R}^{SM})^{2}}\right|\lesssim0.009\;,\quad(95\%\:\mathrm{C.L.})\label{eq:universalitybound-general}
\end{equation}
For instance, in the case of the operator giving rise to only right-handed
couplings we have $g_{L}^{SM}=-\frac{1}{2}+s_{W}^{2}$ , $g_{R}^{SM}=s_{W}^{2}$
, $\delta g_{\ell R}^{NP}=-\frac{1}{2}\kappa_{\ell\ell}$, with $s_{W}^{2}=0.231$
the weak mixing. Then we obtain
\begin{equation}
\kappa_{\ell\ell}\lesssim0.009\,\frac{1-4s_{W}^{2}+8s_{W}^{4}}{4s_{W}^{2}}=0.005\;,\quad(95\%\:\mathrm{C.L.})\,.\label{eq:universalitybound-right}
\end{equation}
Limits on the diagonal elements directly translate into upper limits on the LFV ones, $|\kappa_{\ell \ell^\prime}|<\sqrt{\kappa_{\ell\ell}\,\kappa_{\ell^\prime\ell^\prime}}$, as the $\kappa$ matrix is Hermitian and positive-definite. Thus, together with the constraints on off-diagonal $\tau\mu$ couplings, see eq.~\eqref{derZ}, we will require that
$|\kappa_{\ell\ell^\prime}|\lesssim \mathcal{O}(10^{-3})$, for $\ell,\,\ell^\prime=\mu,\,\tau$. Similar constraints on non-unitarity are obtained in refs.~\cite{FernandezMartinez:2007ms, Antusch:2006vwa,Antusch:2014woa}.

\section{HLFV at one loop} \label{sec:loop}

\begin{figure}
	\centering
        \includegraphics[scale=0.55]{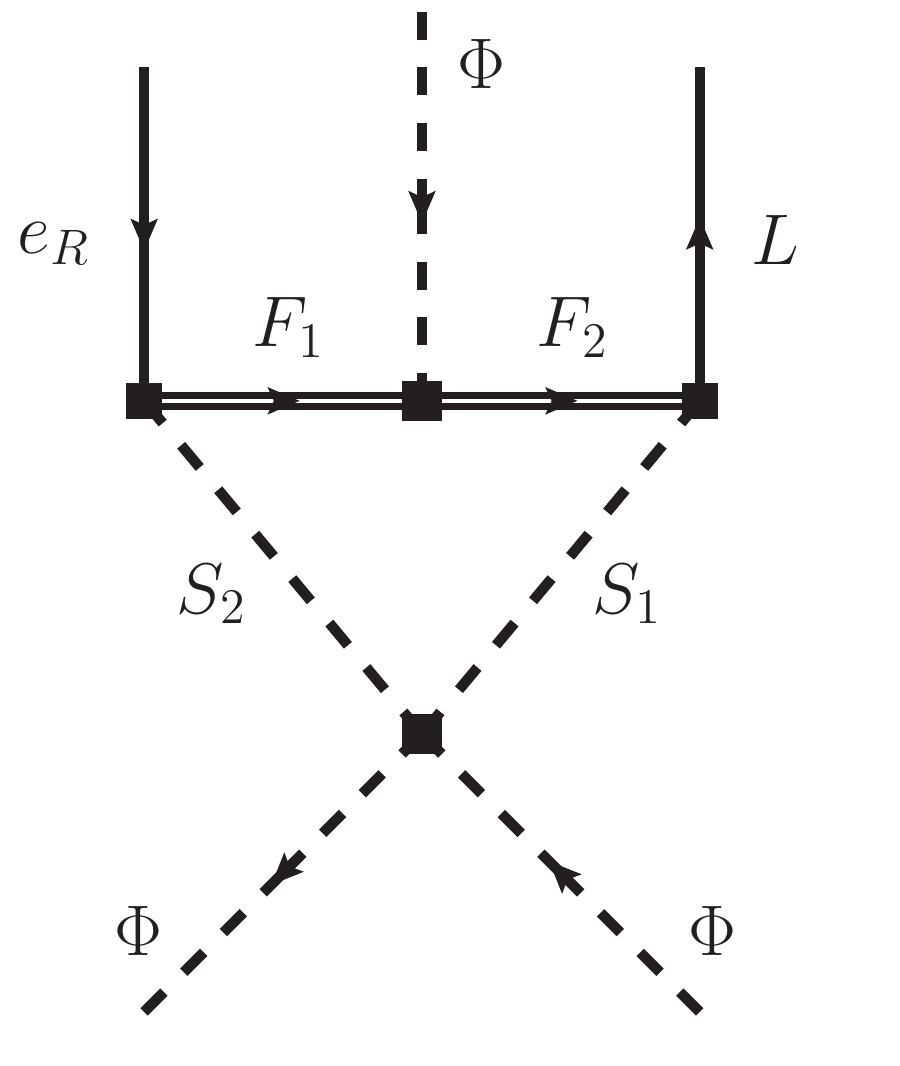}~~~\includegraphics[scale=0.55]{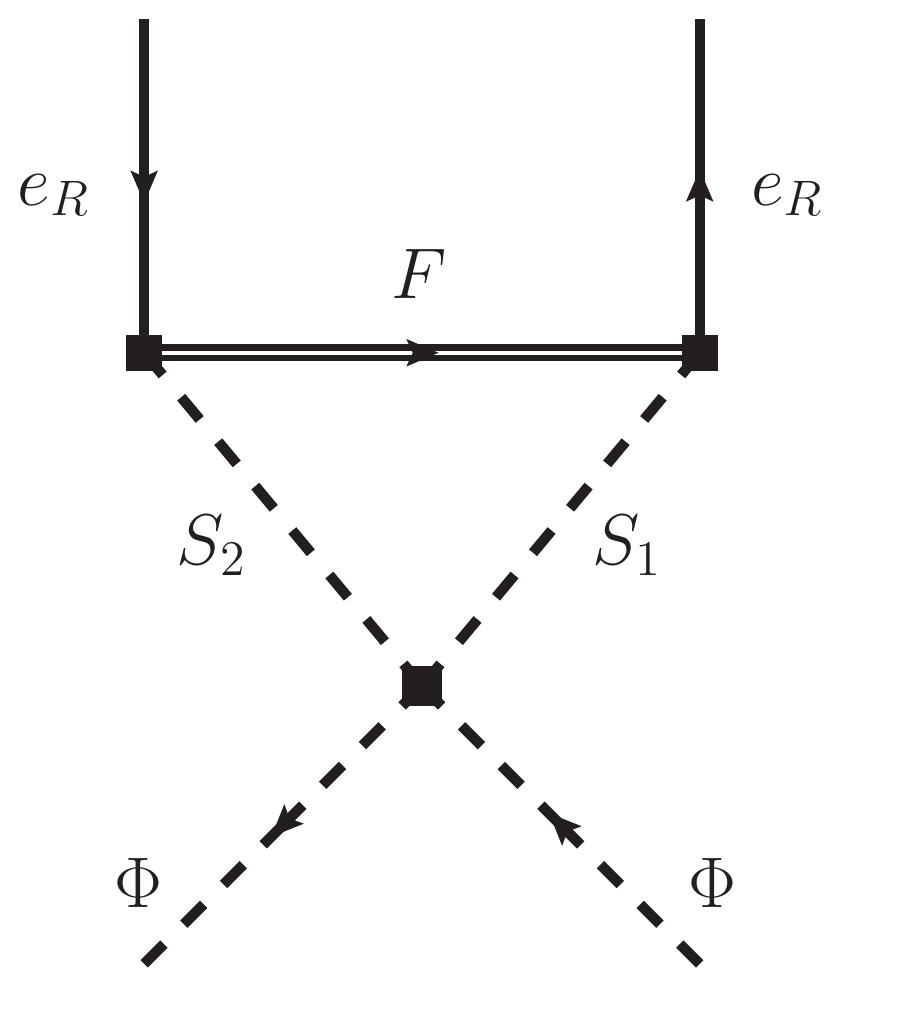}~~~\includegraphics[scale=0.55]{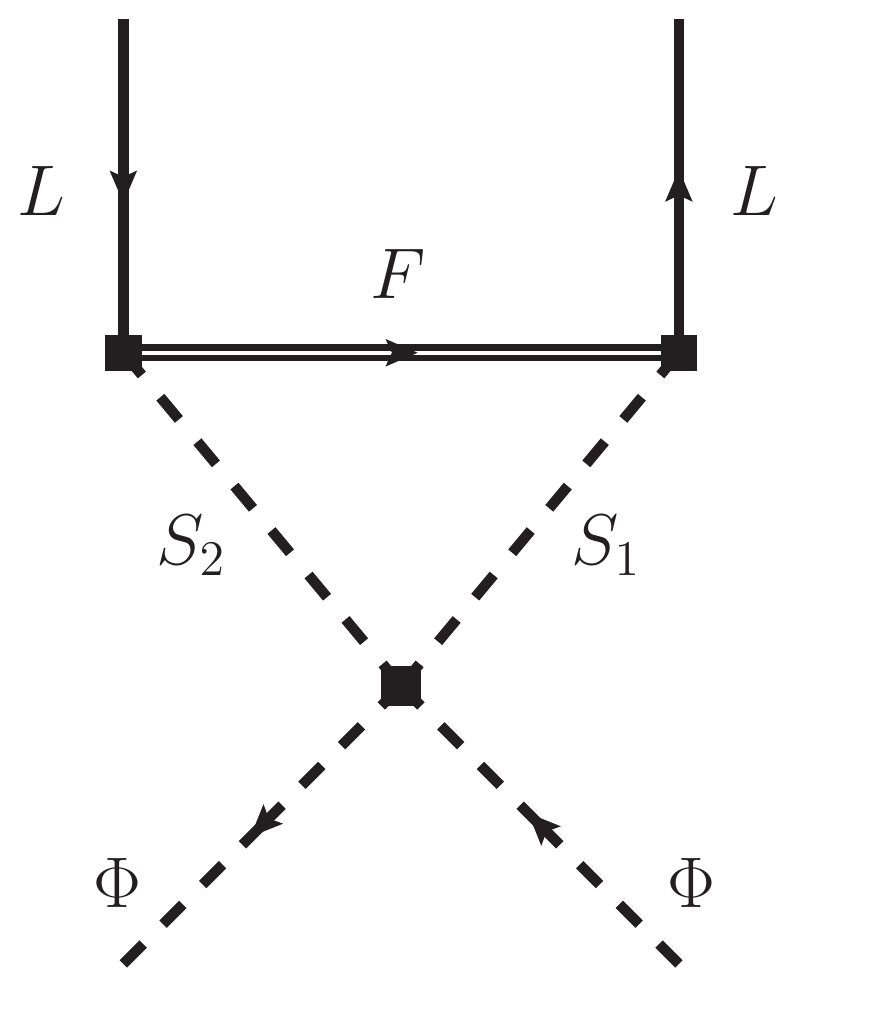} 
	\caption{Different one-loop UV completions of the \emph{Yukawa} operator ${\cal O}_Y$ (left), eq.~\eqref{eq:OH}, and the \emph{derivative} operators ${\cal O}_{1R,\,2R}$ (middle) and ${\cal O}_{1L,\,2L}$ (right), eqs.~\eqref{eq:O1} and \eqref{eq:O2}. In general $F$  will be new fermions and $S$ new scalars, but in some models $F$ can also represent one of the SM leptons.}
\label{openoploop}
\end{figure}

Possible topologies for HLFV at one loop are shown in figure \ref{openoploop}. 
Contrary to the previous tree-level analysis, this is not an exhaustive list but only some representative examples, most of them appearing in 
well-known neutrino mass models that we will discuss in the next section.

One remark about HLFV at one loop is in order: while in the EFT it is very easy to identify the diagrams which generate the relevant operator, most of the calculations have been performed in particular models, where one has to take care properly of one loop (finite) field renormalizations, to enforce flavor diagonal  kinetic terms  at the loop-level.
As a consequence, one needs to check that the result has the correct decoupling behavior in 
the limit $M \gg v$, where $M$ stands for the scale of new physics in the model, that is, 
 the amplitude should scale as $v^2/M^2$, since at the effective Lagrangian level HLFV is mediated 
by the operators in eqs.~\eqref{eq:OH}, \eqref{eq:O1} and \eqref{eq:O2}.

Very roughly we can distinguish two general types of models, which lead to different expectations 
for HLFV at one loop. 
In  models where HLFV is generated via the \emph{derivative} operator (middle and right diagrams in figure~\ref{openoploop}), 
the use of the equations of motions leads to HLFV proportional to the SM lepton 
Yukawa couplings. The diagram on the left of figure~\ref{openoploop} flips helicity in the Yukawa couplings of the new fermions, $F_{1,2}$ and gives rise to the 
\emph{Yukawa} operator. Note, however, that in many neutrino mass models the fermions running in the loop are SM leptons which also give a $m_\tau$ suppression from one of the SM Yukawa couplings. In those cases $C_Y$ is proportional to $y_\tau$, and we can estimate the $h\rightarrow \mu \tau$ rate as
\begin{equation}
\mathrm{BR}(h\rightarrow \mu \tau) \sim 1200 \,y_\tau^2\,\frac{ \lambda_{i H}^2}{(4\pi)^4} \Big(\frac{v}{\rm TeV}\Big)^4\, \Big(\frac{Y}{M_i/\text{TeV}}\Big)^4 \ , 
\end{equation}
where $\lambda_{i H}$ is the quartic coupling of the new scalar to the Higgs, and $Y$
its coupling to leptons (SM and additional ones).
Moreover, some of the new particles in the loop have to be electrically charged, and therefore $\tau \rightarrow \mu \gamma$ will always be present.
Then, bounds from the non observation of $\tau \rightarrow \mu \gamma$, which is usually proportional to $y_\tau^2$, typically give the constraint
\begin{equation}
\Big(\frac{Y}{M_i/\text{TeV}}\Big)^4\lesssim\mathcal{O}(1)\,, 
\end{equation}
which leads to $\mathrm{BR}(h \rightarrow \mu \tau) \lesssim 10^{-8}$ and provides an example of column two and row four in table \ref{tab:CHvsCgamma}.

Notice that in these scenarios
there are also contributions to $h \rightarrow \gamma \gamma$ and $h \rightarrow Z\gamma$. As currently LHC experiments no longer see a significant excess on the $\gamma\gamma$ channel, this imposes further limits on the possibility of having sizable HLFV in these radiative models. 

Even if the CLFV constraints are somehow avoided  with a particular texture 
and/or  some fine-tuned cancellations,  
taking  $Y\sim 1$ and $M_i\sim v$, one still expects a fairly small HLFV. 
For $\lambda_{i H} \sim \mathcal{O}(1)$, the general expectation is:
\begin{equation}
\mathrm{BR}(h \rightarrow \mu \tau) \lesssim 10^{-5} - 10^{-4}  \ ,  
\end{equation}
allowing for some enhancement due to possible numerical factors and couplings of order one, 
at most. This is in agreement with the explicit calculations in the MSSM see-saw  
\cite{Arganda:2004bz}.

More promising scenarios are those in which the chirality flip  in the \emph{Yukawa} operator 
is provided by a new coupling, $Y_{LR}$, since then there is typically  an enhancement factor  
$(Y_{LR}/y_{\tau})^2 \sim 10^4$ for $Y_{LR} = {\cal O}(1)$ (providing examples of row three in table \ref{tab:CHvsCgamma}).
Still, to have sizable contributions one needs particular flavour structures that suppress CLFV and enhance HFLV, and large couplings, $\sim \sqrt{4 \pi}$, which could lead to instabilities or non-perturbativity close to the EW scale. 
This mechanism is at work for instance in SM extensions with scalar leptoquarks coupled to the top quark, studied in refs.~\cite{Dorsner:2015mja, Cheung:2015yga, Baek:2015mea}.

A similar enhancement can be obtained 
in models with both singlet and doublet extra scalar fields (other than type-III 2HDM), 
with slightly smaller dimensionless couplings but large 
trilinear ones, $\mu \Phi^\dagger \phi_\ell \phi_e$. In
 this case the $\mathrm{BR}(h \rightarrow \mu \tau)$ can be further   enhanced by a factor 
$(\mu/M)^2   \sim (5-10)^2$, being $M$ the scale of the new particles in the loop \cite{Baek:2015fma}. However, this trilinear coupling cannot be very large by naturality (and also charge breaking constraints). 

The minimal supersymmetric standard model (MSSM) is  a natural candidate to produce 
HLFV.
In spite of containing two SM scalar doublets, LFV Higgs decays in the MSSM are generated at one loop, because the  holomorphicity of the the superpotential prevents the coupling of the two 
doublets to both, charged leptons and neutrinos. 
LFV Higgs decays  within the MSSM have been studied by several groups, 
finding generically 
a $\mathrm{BR}(h \rightarrow \mu \tau)$ several orders of magnitude below the present 
experimental sensitivity 
both in the R-parity conserving 
\cite{Brignole:2003iv, DiazCruz:2008ry, Arana-Catania:2013xma, Arganda:2015uca}
as well as in the R-parity violating case \cite{Arhrib:2012ax}, as expected 
from our estimates above.
However, a detailed study shows that it is possible to find very fine-tuned regions of the supersymmetric parameter space
 with large $\tan\beta$ and $\mu$-term close to its perturbative bound, 
 in which $\mathrm{BR}(h \rightarrow \mu \tau)$ can reach the percent level 
 satisfying also present limits from $\tau \rightarrow \mu \gamma$ and 
 $h \rightarrow \gamma \gamma$ \cite{Aloni:2015wvn}.
Analogously,  it is possible to obtain such a large ratio in some  
regions of the parameter space of the  supersymmetric inverse see-saw model, 
for particular structures of (very large)  Yukawa couplings, which induce 
LFV slepton masses \cite{Arganda:2015naa}.  

In summary, we conclude that when HLFV occurs at one loop a very suppressed rate is 
expected generically, $\mathrm{BR}(h \rightarrow \mu \tau) \lesssim10^{-4}$, but in most of the models it is much smaller. This is
due to the loop suppression factor, in many cases a $y_\tau$ suppression and finally the CLFV constraints. Although sometimes 
these constraints can be avoided, with a certain amount of fine tuning, 
HLFV at the percent level requires Yukawa couplings or  trilinear couplings
close to their perturbative limit.
In the next section, we also illustrate this generic estimates within the context of several neutrino mass models.

\section{HLFV and neutrino masses} \label{sec:nus}

As we have mentioned before, up to now, the only evidence for LFV are neutrino oscillations, 
which also imply that neutrinos are massive.  Then, since
new physics is needed to explain their mass, 
it is natural to wonder whether  the excess observed by CMS in 
$h\rightarrow \tau \mu$ can be accommodated within any neutrino mass model. It is true that, in general, one expects the new physics scales responsible for neutrino masses to be much above the TeV scale, while for large HLFV one needs new physics at the TeV. However, this does not need to be true in all neutrino models, in particular radiative neutrino mass models and left-right symmetric models can provide neutrino masses with new physics at the TeV scale. 
Obviously, any model containing the type-III 2HDM can potentially explain HLFV, 
with possibly further restrictions due to the neutrino sector.
Interestingly enough, there are  two well-known models for neutrino masses 
which naturally include two  ${\rm SU(2)}$ scalar doublets coupled to leptons: 
the Zee model and the aforementioned left-right (LR) symmetric models, that we will discuss in the following.

On the other hand, in most models of neutrino masses HLFV is not generated at 
tree level but at one loop; we will estimate the expected size of 
$h\rightarrow \tau \mu$  in some of the better motivated models and conclude 
on general grounds that whenever HLFV is generated at one loop the typical size 
is too small to account for the CMS excess.

\subsection{HLFV at tree-level} \label{nus-t}

\subsubsection{The Zee model}

One of the simplest models for neutrino masses is the Zee model~\cite{Zee:1980ai}. It introduces an extra Higgs doublet $\Phi_2$ and a singly-charged singlet $s^+$. In order to explain neutrino mixings correctly it is necessary that both scalar doublets couple to the charged leptons and thus it is a type-III 2HDM, see for instance ref.~\cite{He:2003ih}. In the Zee model, however, it is not clear whether a sizable HLFV can be obtained as there are extra constraints coming from neutrino masses and mixings, as well as new contributions to CLFV from the extra singly-charged scalar. Therefore, it is very interesting to discuss it qualitatively, but a full detailed analysis of the model including HLFV and all CLFV constraints is beyond the scope of this work and is currently in preparation~\cite{us}. Notice that although the model only deals with the lepton sector, there needs to be couplings to quarks of the light Higgs in order for it to be SM-like, as observed. We will assume the simplest scenario in which there is no flavor violation in the quark sector and the interactions are SM-like, such that all production and decays of the lightest Higgs are SM-like.

The most general Yukawa Lagrangian reads:
\begin{equation}
\mathcal{L}_{Y}=
-\overline{L}\, (Y^\dagger_1 \Phi + Y^\dagger_2 \Phi_2)e_{\rm R}  -  \overline{\tilde{L}}f\,L s^{+}+\mathrm{H.c.}  \,, \label{eq:yuk2d}
\end{equation}
where $\tilde{L} \equiv i \sigma_2 L^c = i \sigma_2 C \overline{L}^T$. Due to Fermi statistics, $f_{ab}$ is an antisymmetric matrix in flavour space, while $Y_1, Y_2$ are completely general. The scalar potential is the general 2HDM potential with extra terms coupling the scalar doublets $\Phi,\,\Phi_2$ with the singly charged $s^+$, see e.g. ref.~\cite{Kanemura:2000bq}. For our purposes, the relevant piece in the scalar potential is
\begin{equation}
\mu\,  (\tilde\Phi^T \Phi_2^*)\,s^{+}+\mathrm{H.c.}\,,
\end{equation}
as it violates lepton number by 2 units.

\begin{figure}
	\centering
        \includegraphics[scale=0.55]{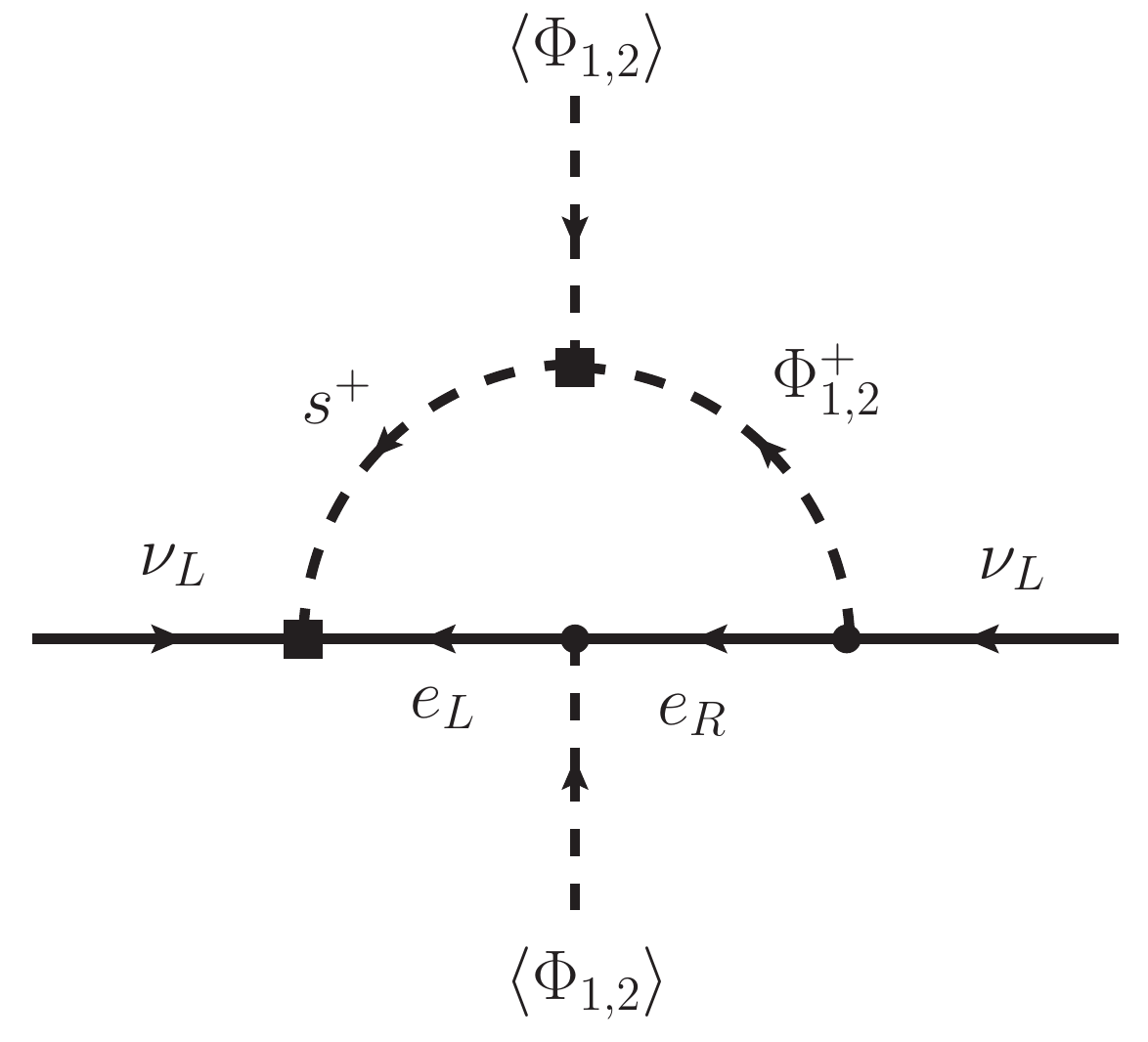}
	\caption{The Zee model diagram for neutrino masses.} \label{Zee}
\end{figure}

Both Higgs doublets take a VEV ($\langle \Phi \rangle=v_1/\sqrt{2}, \,\langle \Phi_2 \rangle=v_2/\sqrt{2}$), and therefore the charged lepton masses are given by (using the short-hand notation $\sin\beta\equiv s_\beta, \,\cos\beta\equiv c_\beta$, $\tan\beta\equiv t_\beta=v_1/v_2$)\footnote{Notice that in this section we have changed the definition of $\tan\beta$.}:
\begin{equation}
m_f=\frac{v}{\sqrt{2}}(s_\beta Y^\dagger_1+c_\beta Y^\dagger_2)\,.
\end{equation}
We will work in the basis where $m_f$ is diagonal. 

Regarding HLFV, $h\rightarrow \mu \tau$ is given by:
\begin{equation}  \label{BRH}
\mathrm{BR}(h\rightarrow \mu \tau) =\frac{m_h}{8\pi\Gamma_h} \,\left(\frac{\,s_{\beta-\alpha}}{\sqrt{2}\,s_\beta}\right)^2\,(|Y_2^{\tau \mu}|^2+|Y_2^{\mu \tau}|^2),
\end{equation}
where $\alpha$ is the angle that rotates the neutral scalars, and similarly for $\mathrm{BR}(h\rightarrow \tau e)\propto (|Y_2^{\tau e}|^2+|Y_2^{e \tau}|^2)$.

Let us now discuss how neutrino masses impose constraints on the  size of the HLFV decays. The diagram shown in fig.~\ref{Zee} generates neutrino masses at one loop. The complete neutrino mass matrix reads (see for instance ref.~\cite{He:2011hs}):
\begin{equation}
M_\nu= A\,\Big(f\,m_f^2+m_f^2f^T-\frac{v}{\sqrt{2}\,c_\beta}(f\,m_f\,Y_2+Y_2^T\,m_f\,f^T)\Big),
\end{equation}
where
\begin{equation}
A\equiv \frac{\sin (2\theta_z)}{8 \sqrt{2}\pi^2\, v\, t_\beta}\,\ln\left(\frac{m^2_{s^+_2}}{m^2_{s^+_1}}\right),
\end{equation}
with $\theta_z$ being the mixing angle for the charged scalars with masses $m_{s^+_1}, \,m_{s^+_2}$, which is proportional to $\mu$, the lepton-number-violating parameter.

The important point is that one needs $Y_2\neq0$ in order to account for neutrino mixing angles. If there are no big hierarchies between $Y_2$ and $f$, neglecting $m_e,m_\mu \ll m_\tau$, one can get the following Majorana mass matrix (which is of course symmetric):
 \begin{equation}  \nonumber
M_\nu =A\times\, \frac{m_\tau\, v}{\sqrt{2}\,c_\beta}\times
 \end{equation}
 \begin{eqnarray}
\times\left(\begin{array}{ccc}
-2 f^{e\tau}\,Y_2^{\tau\,e}\,\,\,\,\,\,\,\,\,\,&-f^{e\tau}\,Y_2^{\tau\,\mu}- f^{\mu\tau}\,Y_2^{\tau\,e}\,\,\,\,\,\,\,\,\,\,& \frac{\sqrt{2}\,c_\beta\,m_\tau}{v}\,f^{e\tau}-f^{e\tau}\,Y_2^{\tau\,\tau}\\
-\,\,\,\,\,&-2 f^{\mu\tau}\,Y_2^{\tau\,\mu}\,\,\,\,\,\,\,\,\,\,&\frac{\sqrt{2}\,c_\beta\,m_\tau}{v}\,f^{\mu\tau}-f^{\mu\tau}\,Y_2^{\tau\,\tau}\\
-\,\,\,\,\,&-\,\,\,\,\,&0\end{array}\right)\,,\label{Mnu}
 \end{eqnarray}
where the dashes refer to the elements $(M_\nu)_{ij} = (M_\nu)_{ji}$ for  $i > j$ and all yukawas can be taken to be real except for $Y_2^{\tau\,\mu}$. In order to have correct mixing angles we need $Y_2^{\tau\mu}$ and $Y_2^{\tau e}$, which enter in the $1-2$ submatrix to be non-zero.  Therefore there are correlations between mixing angles and $\mathrm{BR}(h\rightarrow \mu \tau), \mathrm{BR}(h\rightarrow \tau e)$, and in fact a lower bound on the product of both rates is expected. 

In order to have sizable $h\rightarrow \tau \mu$ we need sizable $Y_2^{\tau \mu}$ and/or $Y_2^{\mu \tau}$.\footnote{Notice that $Y_2^{\mu \tau}$ ($Y_2^{e \tau}$) only enter in the neutrino mass matrix at order $\mathcal{O}(m_\mu)$ ($\mathcal{O}(m_e)$), and is therefore  subject to weaker constraints, at least as neutrino masses are concerned, than $Y_2^{\tau \mu}$ ($Y_2^{\tau e}$). However both enter in CLFV, see below.} This means that it is preferable to have small $f_{e \tau}$ and $f_{\mu \tau}$ to reproduce the correct mixing angles and have a large HLFV, while the overall scale can be adjusted conveniently with the LNV parameter $\mu$ and/or the scalar masses. For typical values of $A\sim 10^{-4}\, \rm GeV^{-1}$, the correct neutrino mass scale can be obtained $Y_2^{\tau \tau},Y_2^{\tau \mu}\gtrsim Y_2^{\tau e}\sim\mathcal{O}(0.01)$ and very small singly-charged couplings, $f_{e\tau},f_{\mu \tau}\sim10^{-8}$ (see for instance~\cite{He:2011hs} for a numerical scan of the model). Moreover, \emph{small} $f_{e \tau}$ and $f_{\mu \tau}$ also suppress the singly-charged contribution to CLFV, which will therefore be dominated in the range of interest for HLFV by the scalars coming from the doublets.

In type-III 2HDM there is an upper bound on $\mathrm{BR}(h\rightarrow \mu \tau)\times \mathrm{BR}(h\rightarrow \tau e)$ from combining the rates of $\mu\rightarrow e\gamma$ and $\mu e$ conversion (which currently saturates the bound). The key point is that all combinations of couplings relevant to these HLFV processes enter in CLFV when the tau-leptons run in the loop. Neglecting the dipole contribution to $\mu e$ conversion, it reads:\footnote{Notice that using EFT the upper bound is much weaker, around $\sim10^{-4}$~\cite{Dorsner:2015mja}.}
\begin{equation}  \label{BR}
\mathrm{BR}(h\rightarrow \mu \tau)\times \mathrm{BR}(h\rightarrow \tau e) \lesssim 10^{-8},
\end{equation}
with a strong dependence on $t_\beta$ and $m_A$, see figure 4 of ref.~\cite{Dorsner:2015mja}\footnote{The authors use an MSSM motivated scenario. It is possible that the constraint can be somewhat weakened by considering a type-III 2HDM with non-related parameters.}. This is clearly a very strong constraint, and the compatibility with correct mixing angles in the Zee model is under investigation~\cite{us}. Indeed, if compatible, future data confirming the $\mathrm{BR}(h\rightarrow \mu \tau)\sim 0.01$ excess would imply an upper bound $\mathrm{BR}(h\rightarrow \tau e) $ at the order of $\sim 10^{-6}$~\cite{Dorsner:2015mja}.  

Furthermore, in the simplest version neglecting the muon mass, one neutrino is massless and only inverted hierarchy is possible. Therefore $0.01\,\mathrm{eV}  \lesssim |m_{ee}| \lesssim 0.05\,\mathrm{eV}$, which can be observed in planned $0\nu2\beta$ experiments. This implies that there is a lower bound on $\mathrm{BR}(h\rightarrow \tau e)$ from the lower bound on $|m_{ee}|>0.01$\,eV. All these correlations make the model phenomenologically very interesting.

We want to conclude this section by emphasizing that whether satisfying all requirements (especially eq.~\eqref{BR}) and having correct mixing angles is possible or not requires a detailed study with a full parameter scan that is beyond the scope of this work~\cite{us}.

\subsubsection{Left-right symmetric models}

Left-right (LR) symmetric models provide a natural realization of the see-saw mechanism 
of type I and II,  together with an elegant explanation for the origin of  parity violation
in the electroweak interactions \cite{Mohapatra:1979ia, Mohapatra:1980yp, Senjanovic:1975rk, Keung:1983uu,Mohapatra:1986uf}. 
The see-saw scale in these models is tied to the scale of
parity breaking, and, if sufficiently low, they can be tested at the LHC as well as in low energy experiments such as neutrinoless double beta decay and CLFV.

The minimal LR see-saw model is based on the gauge group 
$\rm SU(2)_L\times SU(2)_R\times U(1)_{B-L}$, with equal left and right gauge couplings, 
$g_{\rm L}=g_{\rm R}$. 
 The fermions are completely left-right symmetric, in particular in the lepton sector 
 $L_L=(\nu_L\,e_L)^T$ ($L_R=(\nu_R\,e_R)^T$) is a doublet under the $\rm SU(2)_L$ ($\rm SU(2)_R$) group.
The electric charge is obtained from the well-known relation
\begin{equation}
Q=T_{3L}+T_{3R}+\frac{B-L}{2}\,,
\end{equation}
where $\rm T_{3L,3R}$ refer to the third component of weak isospin of $\rm SU(2)_{L,R}$.

The scalar sector consists of a bi-doublet $(2,2,0)$ which is added in order to give masses to the quarks and leptons,
\begin{eqnarray}
\Sigma=\left(\begin{array}{cc} \Phi_1^0\,\, &\Phi_2^+\\
\Phi_1^-\,\, &\Phi_2^0 \end{array}\right) \ ,
\qquad\qquad \tilde\Sigma=\sigma_2\,\Sigma^*\,\sigma_2=\left(\begin{array}{ccc} \Phi_2^{0*}\,\, &-\Phi_1^+\\
-\Phi_2^-\,\, &\Phi_1^{0*} \end{array}\right) \ , 
\end{eqnarray}
and two triplets, $\Delta_{\mathrm{R}}(1,3,2)$
and $\Delta_{\mathrm{L}}(3,1,2)$, according to their 
$\rm SU(2)_L\times SU(2)_R\times U(1)_{B-L}$ quantum numbers. 
In a first stage the gauge symmetry $\rm SU(2)_R\times U(1)_{B-L}$ is broken by the VEV
$\langle \Delta^0_R \rangle = v_R$ down to the $U(1)_Y$ of the SM. In the second stage, 
the neutral components of $\Sigma$ develop a VEV and break the SM symmetry down to 
$U(1)_{\rm em}$, 
\begin{equation}
\langle \Sigma \rangle =\frac{1}{\sqrt{2}} \left(\begin{array}{cc} v_1\,\, & 0 \\
0 \,\, &v_2 e^{i\alpha} \end{array}\right) \ ,
\end{equation}
where $v_{1,2}$ are real and positive, and in the following we take $\alpha=0$ for simplicity, since it does not affect our discussion. The minima of the potential yield the relation $v_L\cdot v_R \sim \gamma \,v^2$, where 
$v^2=v_1^2+v_2^2$, $v_L$ is the (tiny) VEV of $\Delta_L$  and $\gamma$ is some combination of the 
scalar potential parameters and the bi-doublet VEVs, $v_1,v_2$ (see refs.~\cite{Mohapatra:1980yp, Deshpande:1990ip} for more details).

The new gauge bosons $Z_{\rm R}$ and $W_{\rm R}$ obtain their masses primarily from $v_{\rm R}$,  
while $Z,\,W_{\rm L}$ obtain their masses from the VEVs $v_1,v_2$ of the bi-doublet (and from 
$v_{\rm L}\ll v_1,v_2$).  
We neglect the mixing among left and right gauge bosons, which comes from the product of the VEVs $v_1$ and $v_2$. This is justified by the large $v_{\rm R}$ scale implied by present bounds,
 $v_{\rm R} >  3$ TeV~\cite{Maiezza:2010ic}, which does not require anymore  a small ratio $v_1/v_2$   to suppress $W_{\rm L} - W_{\rm R}$ mixing.

The Yukawa Lagrangian relevant for the leptons reads:
\begin{equation}
\mathcal{L}_{Y}=
-\overline{L}_L\, (Y_1 \Sigma + Y_2 \tilde\Sigma)L_{\rm R} \,+\rm\,H.c.\,, \label{eq:yukLR}
\end{equation}
where $Y_1, Y_2$ are completely general. Similar interactions exist in the quark sector, with the appropriate Yukawa couplings. If one imposes the natural left-right symmetry $L_L \rightarrow R_R,\, \Delta_L\rightarrow \Delta_R,\, \Sigma\rightarrow \Sigma^\dagger$, see for instance ref.~\cite{Deshpande:1990ip}, then $Y_{1,2}$ are Hermitian matrices, which further restricts the parameter space with respect to a general type-III 2HDM.

After the neutral components of $\Sigma$ take a VEV, we get:
\begin{align}
m_{\mathrm{E}}=(Y_2 c_\beta+ Y_1 s_\beta) v/\sqrt{2}\,,\qquad m_{\mathrm{D}}=(Y_1 c_\beta+ Y_2 s_\beta) v/\sqrt{2}\,,
\end{align}
with $t_\beta=v_2/v_1$. Thus, $Y_{1,\,2}$ can be completely expressed in terms of $m_{\mathrm{E}},\,m_{\mathrm{D}}$. We can work in the basis where $m_E$ is diagonal. The Yukawa couplings of the charged leptons are those of a type-III 2HDM:
\begin{align}
\mathcal{L}_{\rm C} &=\overline{e_{\rm L}}\, \left(Y_1\,\Phi_2^0 + Y_2 \,\Phi_1^{0*}\right)\,e_{\rm R} \,+\rm H.c.\\
  &\approx \frac{1}{\sqrt{2}\,v}\, \overline{e_L}\, \left(m_E + \frac{m_{\mathrm{D}}-2\,s_\beta\,c_\beta\,m_{\mathrm{E}}}{c_{2\beta}}\,\epsilon \right)\, h\,e_{\rm R} \,+\rm H.c. + ...\,, \label{eq:yukC}
\end{align}
 where we have first rotated $\Phi_{1}^0 \approx 1/\sqrt{2}\,(c_\beta h^\prime - s_\beta H^\prime)$ and $\Phi_{2}^0 \approx 1/\sqrt{2}\,(c_\beta H^\prime + s_\beta h^\prime)$, and then used that $H^\prime\approx H + \epsilon \,h$, with the mixing being $\epsilon \propto v^2/v_R^2$ \cite{Deshpande:1990ip}. Notice thus, that the LFV Higgs decays will be controlled by $m_D$ and suppressed by the mixing squared, roughly $\epsilon^2 \propto v^4/v_R^4$, which phenomenologically is $\lesssim 10^{-4}$. This suppression can be partially overcome by taking the quartic couplings of the scalar potential which enter in the mixing large enough.

For $\tan \beta\sim \mathcal{O}(1)$, one expects Dirac masses for neutrinos of the same order
as the charged leptons. 
However, small (large) Majorana masses of the LH (RH) neutrinos are also present from the VEV of the triplet, $v_{\rm L}$ ($v_{\rm R}$). Therefore, the large Majorana mass of the RH neutrinos yield light  neutrino masses via see-saw type I , plus a see-saw type II contribution~\cite{Mohapatra:1980yp}:
\begin{equation}
m_\nu=f\, v_{\mathrm{L}}- m_{\mathrm{D}}\,(f\,v_{\mathrm{R}})^{-1}\,m_{\mathrm{D}}^T \,,
\end{equation}
where $m_{\mathrm{D}}$ encodes the connection between neutrino masses and HLFV. Notice that for $v_R$ to be at the TeV scale, the Dirac neutrino masses 
can not be larger that ${\cal O}({\rm MeV})$, which may 
 require some fine tuning, due to the relation with charged lepton masses. Furthermore, 
 we need $\gamma \ll1$ in order to have a small enough type-II see-saw contribution to neutrino masses.

Regarding the bi-doublet, one of the $\rm SU(2)_L$ doublet gets a large mass, proportional to 
$v_R$, while the other remains light, at the weak scale.
Thus the phenomenology of this  sector is similar to the type-III 2HDM
close to the alignment limit. However, in this minimal model the strong bounds from neutral flavour violating Higgs (FCNH) interactions in the quark sector imply that 
the second Higgs mass should be $m_H \gtrsim 15 $ TeV~\cite{Beall:1981ze, Zhang:2007da, Maiezza:2010ic, Bertolini:2014sua}, leading to 
$\mathrm{BR}(h\rightarrow \tau \mu)$  that is too small. 
Therefore in order to get HLFV observable at LHC the scalar sector has to be extended, 
in such a way to avoid the bounds from FCNH effects in the quark sector while keeping 
them at LHC reach in the lepton sector. 
One possibility is just to assume a different pattern of interactions with the scalars  
in each sector, for instance the minimal one in the lepton sector, which leads to the 
type-III 2HDM, but not in the quark sector.  
Alternatively, one can consider two (or more) bi-doublets and particular Yukawa structures which 
avoid the undesired quark flavor-changing effects.\footnote {Two bi-doublets fit well into the 
supersymmetric version of the minimal LR model.}

Assuming that the scalar sector is modified so as to allow for a second SM doublet 
light enough,  in order to get a sizable $\mathrm{BR}(h\rightarrow \tau \mu)$ we still need large 
$Y_2^{\tau \mu}$, which may be in conflict with CLFV constraints, since new contributions from 
$W_{\rm R}$ and $Z^\prime$ are present besides those of a generic type-III 2HDM.
A more detailed study of HLFV in the context of a LR symmetric scenario with an extended scalar sector and its viability is beyond the scope of this work.

\subsection{Neutrino mass models with HLFV at one loop} \label{loop}

We will now consider some well-known models of neutrino masses in which HLFV 
appears at one loop.
The possible topologies within such neutrino mass models 
 both for the \emph{Yukawa} operator, eq.~\eqref{eq:OH}, 
  and the \emph{derivative} operators, eqs.~\eqref{eq:O1} and~\eqref{eq:O2},
   are shown in figure~\ref{openoploop}:
left-right external fermion legs corresponds to the \emph{Yukawa} operator (left figure), and right-right (middle) and left-left (right) correspond to the \emph{derivative} operators.
Depending on the chirality of the fermions in the fermion-fermion-scalar vertex of the loop,
 we have  the different neutrino mass  models, see table \ref{tabletoploop}.
 Also, the scalar $S$ in the loop denotes both, the SM Higgs doublet 
 (in see-saw I and III), as well as the extra scalars in models with 
 an extended scalar sector. Similarly, the fermions $F$ may refer to 
 SM leptons and also to the extra leptons specified in table \ref{tabletoploop}. 

Next, we estimate the HLFV rates of these neutrino-related scenarios.

\begin{table}[t]
\centering
\begin{tabular}{ | c | c |  c |c |}
\hline
Topology & Particles & Representations & Neutrino mass model\\ \hline \hline 
${\cal O}_Y$ &  F &  $(1,0)_F, (3,0)_F$ & Dirac, SSI, SSIII (ISS)\\ \hline 
${\cal O}_{1R,2R}$ &S & $(1,2)_S$ & ZB (doubly-charged)\\ \hline
${\cal O}_{1L,2L}$ & S &$(1,1)_S, (3,1)_S$& ZB (singly-charged),  SSII \\
\hline
${\cal O}_{1L,2L}$ (with $Z_2$) &S,F & $(2,1/2)_S$  $\oplus$ $(1,0)_F, (3,0)_F$ & Scotogenic model (IDM) \\ \hline
\end{tabular}
\caption{Particle content of popular neutrino mass models with one-loop topologies, see figure~\ref{openoploop}.} \label{tabletoploop}
\end{table}

\subsubsection{Dirac neutrinos, see-saw type I and III, inverse see-saw}

A thorough  computation of HLFV in both,  type-I see-saw (including the Dirac limit)
and MSSM-see-saw, has been performed 
in ref.~\cite{Arganda:2004bz}.  In the following we just estimate the expected ratios in 
the non-SUSY scenario (SUSY contributions, for the reasons explained in previous section, are also typically below present experimental sensitivity), and refer the reader to  ref.~\cite{Arganda:2004bz} for details. 

If the SM is just extended to include Dirac neutrino masses, the effective Lagrangian approach is not appropriate because the new particles (right-handed neutrinos) are necessarily light. Still, diagrams like those in figure  \ref{openoploop} with neutrinos and charged Goldstone bosons running in the loop can be used to estimate the $h \rightarrow \tau \mu$ amplitude in the gaugeless limit ($g\rightarrow 0$): 
\begin{equation}
A\sim \frac{ m_\tau (Y Y^\dagger)_{\mu\tau}}{(4\pi)^2 v} =
\frac{ m_\tau (m  m^\dagger)_{\mu\tau}}{(4\pi)^2 v^3} =
\frac{ m_\tau}{(4\pi)^2 v^3} \Big(\sum_i U_{\mu i} m_{\nu_i}^2 U^*_{\tau i}\Big)  \ ,  
\end{equation}
with $Y$ the neutrino Yukawa coupling, $m$ the neutrino mass matrix, $m_{\nu_i}$ the neutrino masses and $U_{\alpha i}$ the neutrino mixing matrix. We also assumed the scalar quartic coupling $\lambda\sim 1$. For Dirac neutrino  masses  $m_{\alpha i}\sim 5\times 10^{-11} \text{GeV}\sim 10^{-13}\,v$ this yields negligible $\mathrm{BR}(h \rightarrow \mu \tau)  \sim 10^{-56}$. 

In a complete calculation in the unitary gauge, this result is obtained because the unitarity of the mixing matrix $U$ produces an exact cancellation unless neutrinos are massive (GIM cancellation). One can be more general and assume that just light neutrinos run in the loop but that the mixing matrix is not unitary. This is the case of many neutrino mass models. In order to impose the bounds on a non-unitary lepton mixing matrix $N$, in particular on the $\mu \tau$ element, we can use the results of refs.~\cite{FernandezMartinez:2007ms, Antusch:2006vwa,Antusch:2014woa} on $N\equiv (1+\eta)U$:
\begin{equation} \label{nonuni}
|N N^\dagger|_{\mu \tau}<0.01\,, \qquad |\eta|_{\mu \tau}<0.005\,.
\end{equation}
In this case there is no GIM-cancellation and the rate is just the one in ref.~\cite{Arganda:2004bz} substituting $U$ by $N$.
We estimate:
\begin{equation}
A\sim \frac{ m_\tau}{(4\pi)^2 v}\Big(\sum_i N_{\mu i} N_{\tau i}^*\Big)  \ , 
\end{equation}
and so:
\begin{equation}
\mathrm{BR}(h \rightarrow \mu \tau) \sim 10^{-6} |\sum_i N_{\mu i} N_{\tau i}^*|^2\lesssim 10^{-10}\,,
\end{equation}
where in the last line we used eq.~\eqref{nonuni}.
Notice that this diagram is not suppressed by light neutrino masses, but instead by the loop factor and by the almost unitarity of the mixing matrix (or constraints on the off-diagonal elements). This is the same dominant contribution, that does not depend on the light neutrino masses (having the mass insertion on an external leg, $m_\tau$), that gives CLFV \cite{FernandezMartinez:2007ms}. CLFV constraints on non-unitarity are very strong, as the constant term (independent of the neutrino masses) in the loop function no longer cancels. The difference of course stems from experimental constraints: CLFV bounds are below $10^{-8}$, while HLFV experimental sensitivity is currently at the level of $\mathcal{O}(0.01)$.

For Majorana neutrinos, assuming three right-handed neutrinos of mass $m_{Ri} > m_h$, one can also estimate that in the gauge limit 
\begin{equation}  \label{SSI}
A\sim \frac{ m_\tau}{(4\pi)^2 v}\Big(\sum_i \frac{m_{\mu i} m_{\tau i}^*}{m_{Ri}^2}\Big) \sim  \frac{ m_\tau}{(4\pi)^2 v}\frac{m_{\nu} }{m_{R}} \ , 
\end{equation}
where in the last step we have assumed a common right-handed mass 
and common Dirac mass, $m_R\gg m_{\alpha i}\equiv m_{D}$, so the light neutrino masses are given by the typical see-saw formula, $m_{\nu}\sim m_{D}^2/m_R \sim 10^{-11} \text{GeV}$.
In this case, 
$\mathrm{BR}(h \rightarrow \mu \tau) \lesssim 10^{-31}$.

In the inverse see-saw scenario, the Yukawa couplings can be much larger than 
the naive see-saw scaling, $Y \sim \sqrt{m_\nu m_R}/v  \sim \mathcal{O}(10^{-6})$, because 
$m_\nu \sim \mu\, m_D^2/ m_R^2$, where $\mu$ is the mass splitting between two pseudo-Dirac
sterile neutrinos with masses $m_R \gg \mu$, 
i.e, LFV and LNV are decoupled. 
The estimate for HLFV is exactly the same as in the left part of eq.~\eqref{SSI}, before using 
the see-saw formula. 
In principle $Y$ could be order one, but there are strong constraints from 
CLFV:
\begin{equation}
\Big(\frac{Y}{m_R/\text{TeV}}\Big)^4< \mathcal{O}(0.1)  \ .
\end{equation}
Therefore, we obtain 
\begin{equation}
\mathrm{BR}(h \rightarrow \mu \tau) \lesssim10^{-10} \,,
\end{equation}
in  agreement with the full one-loop computation \cite{Arganda:2014dta}. Assuming particular Yukawa textures, one could evade CLFV constraints yielding at most 
$\mathrm{BR}(h \rightarrow \mu \tau) \lesssim10^{-5}$ \cite{Arganda:2014dta}, 
for very large Yukawa couplings, $Y \sim 4$, close to the perturbative limit, 
still unobservable.

So we conclude that the predicted HLFV rates are always well below experimental sensitivity for 
Dirac neutrinos and the different see-saw scenarios.

\subsubsection{The Zee-Babu model}
In the Zee-Babu model, there are two extra scalars, one 
singly-charged $s$ and one doubly-charged $k$, coupled to the SM leptons as 
 $\overline{\tilde{L}}fL s^{+}+\overline{e^{c}}g\, e\, k^{++}+\mathrm{H.c.}$.
 The $k^{++}$ ($s^+$) contributions arise from the \emph{derivative} operators of type ${\cal O}_{1R,2R}$ (${\cal O}_{1L,2L}$)
through diagrams middle (right) in figure \ref{openoploop}. 
 As the new scalars couple to different  chirality fields, both
 contributions do not interfere
  and, neglecting the light neutrino masses, we can estimate the contributions to  
  $h \rightarrow \mu \tau$ as 
\begin{eqnarray}
y_{\mu\tau} &\sim &\frac{m_\tau v}{(4\pi)^2}
\frac{\lambda_{s\Phi}}{m_s^2}\,4\, (f_{e\mu}^{*}f_{e\tau}) \ , 
\\
y_{\tau\mu}&\sim& \frac{m_\tau v}{(4\pi)^2} \frac{\lambda_{k\Phi}}{m_k^2}\,4\,
(g_{e\mu}^{*}g_{e\tau}+g_{\mu\mu}^{*}g_{\mu\tau}+g_{\mu\tau}^{*}g_{\tau\tau}) \ , 
\end{eqnarray}
where the Higgs couplings to both new scalars are $\lambda_{s\Phi}|s|^{2}\Phi^{\dagger}\Phi+\lambda_{k\Phi}|k|^{2}\Phi^{\dagger}\Phi$ and the factor of 4 comes from the Feynman rules~\cite{Nebot:2007bc}.

Then, for the branching ratio we find:
\begin{equation}
\mathrm{BR}(h \rightarrow \mu \tau) \sim 1200\,y_\tau^2\,16
\Big(
\frac{ \lambda_{s\Phi}^2 v^4}{(4\pi m_s)^4}\, |f_{e\mu}^{*}f_{e\tau}|^2 +\frac{ \lambda_{k\Phi}^2 v^4}{(4\pi m_k)^4}\, |g_{e\mu}^{*}g_{e\tau}+g_{\mu\mu}^{*}g_{\mu\tau}+g_{\mu\tau}^{*}g_{\tau\tau}|^2
\Big) \ .
\end{equation}

Again, we have to take into account all limits from CLFV and universality to constrain the couplings:

\begin{itemize}
\item
The doubly-charged scalar, $k^{++}$, mediates tree-level LFV decays 
$\ell_a^- \rightarrow \ell_b^+ \ell_c^- \ell_d^-$, which provide the strongest limits on all the 
couplings $g_{ab}$; for instance, 
 $\mathrm{BR}(\tau^{-}\rightarrow\mu^{+}\mu^{-}\mu^{-})<2.1\times10^{-8}$ 
($\mathrm{BR}(\tau^{-}\rightarrow e^{+} e^{-}\mu^{-})<1.8\times10^{-8}$ )
\cite{Beringer:1900zz}
 implies
$|g_{\mu\tau}g_{\mu\mu}^{*}|<0.008\,\left(\frac{m_{k}}{\mathrm{TeV}} \right)^{2}$, and 
($|g_{e\mu}^{*}g_{e\tau}|<0.005\,\left(\frac{m_{k}}{\mathrm{TeV}} \right)^{2}$).
Notice that $g_{\tau\tau}$ can not be bounded by this type of processes, although
it is expected to be of order  
$\sim g_{\mu\mu}\,m^2_\mu/m^2_\tau$  to reproduce neutrino mixings.

\item The singly-charged scalar gives rise to violations of universality. In particular, from muon decay and using the limits of the unitarity of the CKM, one gets $|f_{e\mu}|^{2}<0.007\,\left(\frac{m_{h}}{\mathrm{TeV}} \right)^{2}$~\cite{Herrero-Garcia:2014hfa}. Comparing decays into different charged lepton channels, one obtains the limit $||f_{e\tau}|^{2}-|f_{e\mu}|^{2}|<0.035\,\left(\frac{m_{h}}{\mathrm{TeV}} \right)^{2}$~\cite{Herrero-Garcia:2014hfa}.
	
\item 
Radiative LFV decays $\ell_a^- \rightarrow \ell_b^- \gamma$ are generated at one loop
by both $s^+$ and $k^{++}$, so that  
 $\mathrm{BR}(\tau\rightarrow\mu\gamma)<4.4\times10^{-8}$ \cite{Beringer:1900zz} implies
$\frac{|f_{e\mu}^{*}f_{e\tau}|^{2}}{(\frac{m_{s}}{\mathrm{TeV}})^4}+16 \frac{|g_{e\mu}^{*}g_{e\tau}+g_{\mu\mu}^{*}g_{\mu\tau}+g_{\mu\tau}^{*}g_{\tau\tau}|^{2}}{(\frac{m_{k}}{\mathrm{TeV}})^4}<0.7$. 
where the factor of 16 comes from electric charge.
\end{itemize}

Combining all the bounds, we obtain:
\begin{equation}
\mathrm{BR}(h \rightarrow \mu \tau)_{\rm ZB} \lesssim 10^{-10} \, \lambda_{s\Phi}^2 \ , 
\end{equation}
which for  $\lambda_{s\Phi} \lesssim \mathcal{O}(1)$ (to have perturbativity and stability up to some scale not far from the EW), is at most $\lesssim 10^{-9}$.

\subsubsection{See-saw type II}

Similarly to the ZB model, both singly and doubly charged scalars of the $\rm SU(2)$ triplet 
$\Delta$ with hypercharge $Y=1$ contribute to CLFV and HLFV, although in this case they only couple to 
left-handed leptons, generating just the operator of type ${\cal O}_{1L,2L}$. 
We therefore expect a similar result for $\mathrm{BR}(h \rightarrow \mu \tau)$.

By writing the triplet as a $2 \times 2$ matrix:
\begin{equation}
\Delta =\,\begin{pmatrix}
			\Delta^{+}/\sqrt{2} & \Delta^{++} \\
			\Delta_{0} & - \Delta^{+} /\sqrt{2}
		\end{pmatrix} \, ,
\end{equation}
the relevant Yukawa interaction is:
\begin{equation}  \label{Lst}
\mathcal{L}_{\Delta} =  \left( (g^\dagger)_{\alpha \beta} \, \overline{\tilde L}_\alpha 
		\Delta L_\beta + \mathrm{H. c.} \right)\, ,
\end{equation}
where $g$ is a symmetric matrix in flavour space. 

The potential terms that couple the triplet to the Higgs are:
\begin{equation}
V (\phi, \Delta)\subset \lambda_{\Delta \Phi} \,\phi^{\dagger}\,\Delta^\dagger\, \Delta \phi
			+\left( \mu \,\tilde \phi^{\dagger} \Delta^{\dagger} \phi 
			+ \mathrm{H.c.} \right) + \ldots
\label{eq:triplet-pot}
\end{equation}
Thus the relevant terms for HLFV are:
\begin{align}
\lambda_{\Delta \Phi}\, \phi^{\dagger}\,\Delta^\dagger\, \Delta \phi&\subset \frac{1}{2} \lambda_{\Delta \Phi} v\, h\,(\Delta^{+}\Delta^{-}+2\, \Delta^{++}\Delta^{--})\,,\\
\left( g^\dagger \, \overline{\tilde L} \Delta L + \mathrm{H. c.} \right) &\subset 
 \frac{g^\dagger}{\sqrt{2}} \,\left(\Delta^{+}\,(\overline{\nu^c}\,e+\overline{e^c}\,\nu)+\sqrt{2}\,\Delta^{++}\overline{e^c}\,e)\right) + \mathrm{H. c.} \,.
 \label{eq:triplet}
\end{align}

The main contribution comes from the diagram on the right in figure~\ref{openoploop}. For  similar masses $m_{\Delta^{++}}\sim m_{\Delta^{+}}\equiv m_{\Delta}$, which is a reasonable assumption as they belong to the same multiplet, and neglecting factors of two, we find:
\begin{equation}
\mathrm{BR}(h \rightarrow \mu \tau) \sim 1200 \,y_\tau^2 \frac{ \lambda_{\Delta \Phi}^2 v^4}{(4\pi m_\Delta)^4}\, |g_{e\mu}^{*}g_{e\tau}+g_{\mu\mu}^{*}g_{\mu\tau}+g_{\mu\tau}^{*}g_{\tau\tau}|^2\,.
\end{equation}

We can use CLFV data to constraint the couplings. From \cite{Akeroyd:2009nu}, $\mathrm{BR}(\tau\rightarrow\mu\gamma)<4.4\times10^{-8}$  implies
\begin{equation}
\label{eq:ssii}
\frac{|g_{e\mu}^{*}g_{e\tau}+g_{\mu\mu}^{*}g_{\mu\tau}+g_{\mu\tau}^{*}g_{\tau\tau}|^2}{(\frac{m_{\Delta}}{\mathrm{TeV}})^4}<0.04\,.
\end{equation}
Much as in the ZB model, tree level CLFV mediated by $\Delta_{++}$ 
leads to stronger constraints:
  $\mathrm{BR}(\tau^{-}\rightarrow\mu^{+}\mu^{-}\mu^{-})<2.1\times10^{-8}$, implies
$|g_{\mu\tau}g_{\mu\mu}^{*}|^2/\left(\frac{m_{k}}{\mathrm{TeV}} \right)^{4}<10^{-5}$, and there is  a similar bound for the  combination $g_{e\mu}^{*}g_{e\tau}$. However $g_{\mu\tau}^{*}g_{\tau\tau}$ is only constrained by 
$\mathrm{BR}(\tau\rightarrow\mu\gamma)$, so using the  
 upper bound in eq.~\eqref{eq:ssii} we obtain 
\begin{equation}
\mathrm{BR}(h \rightarrow \mu \tau)_{\rm SSII} < 10^{-10} \lambda_{\Delta \Phi}^2 \ .
\end{equation}
The bound is similar to the ZB model one, due to the presence in both cases of the doubly charged scalars.

\subsubsection{The Scotogenic model}
In the Scotogenic model of neutrino masses (also sometimes referred to as Inert Doublet model (IDM)), the SM is extended by three singlet fermions $N_i$ and one scalar doublet, $\eta$, 
which are odd under a $Z_2$ symmetry, while all SM particles are even~\cite{Ma:2006km}. 
The lightest of the scalars and the $N_i$ is a dark matter candidate. Under the assumption that the scalar doublet $\eta$ does not acquire a VEV, 
neutrino masses 
are generated at one-loop, which allows for a new physics scale of order $\lesssim$ TeV with 
much larger Yukawa couplings than in the see-saw scenarios, due to the loop 
suppression. With respect to HLFV, there is only the contribution coming from the second diagram in figure \ref{openoploop},
which generates the \emph{derivative} operator of type ${\cal O}_{1L,2L}$.

The Higgs coupling to the inert doublet scalar $\eta$ is given by 
\begin{equation}
\lambda_{3}(\eta^\dagger \eta) (\Phi^{\dagger}\Phi)
+ \lambda_{4}(\eta^\dagger \Phi) (\Phi^{\dagger} \eta) + \frac 1 2 \lambda_5
  \left[ (\Phi^\dagger \eta)^2 + \mathrm{H.c.} \right] \,.
\end{equation}
 Only the charged scalars contribute via the $\lambda_{3}$ coupling. Notice that HLFV is 
  unconstrained by neutrino masses, which are proportional only to $\lambda_5$.\footnote{Of course $\lambda_{3}$ enters in the physical masses of the new neutral and charged scalars. The neutral scalars enter in the neutrino mass expression, while the charged ones give rise to CLFV.} 
  
  We can estimate the amplitude of $h \rightarrow \mu \tau$ from the first diagram of figure~\ref{openoploop} to be:
\begin{equation}
y_{\mu\tau} \sim \frac{m_\tau v}{(4\pi)^2}\Big(\frac{\lambda_{3}}{m_0^2} \sum_{i} Y_{\tau i}\, Y_{\mu i}^{*}\Big)\,,
\end{equation}
where $m_0= \max(m_{\eta_c},m_{\mathrm{R}i})$, being  $m_{\eta_c}$ the charged scalar mass 
and $m_{\mathrm{R}i}$ the mass of the $N_i$ heavy neutrino. This leads to the branching ratio 
\begin{equation}
\mathrm{BR}(h \rightarrow \mu \tau) \sim 1200\, y_\tau^2 \, \frac{ \lambda_{3}^2}{(4\pi)^4}\, \Big(\frac{v}{\rm TeV}\Big)^4 \frac{\Big | \sum_{i} Y_{\tau i}\, Y_{\mu i}^{*}\Big |^2}{(m_0/ \text{TeV})^4} \,.
\end{equation}
On the other hand, the branching ratio for $\tau\rightarrow \mu \gamma$
 is given by (we assume $m_{\eta_c} > m_{\mathrm{R}i}$, but a similar result holds in the opposite case):
\begin{equation}
\label{eq:idm}
B(\tau\rightarrow \mu \gamma)=\cfrac{3\alpha}{64\pi G_{F}^{2}}\Big | \sum_{i}\cfrac{Y_{\tau i}\, Y_{\mu i}^{*}}{m_{0}^{2}}f\left[\cfrac{m_{\mathrm{R}i}^{2}}{m_{0}^{2}}\right]\Big |^{2}
\times 0.17\,,
\end{equation}
where $f(x)=\cfrac{1-6x+3x^{2}+2x^{3}-6x^{2}\ln(x)}{6(1-x)^{4}}$, which varies between $1/12$ ($x=1$) and $1/6$ ($x=0$).

$\mathrm{BR}(\tau\rightarrow\mu\gamma)<4.4\times10^{-8}$~\cite{Beringer:1900zz} implies (we take the loop function $f\sim 0.1$, ~the most conservative case):
\begin{equation}
\Big |\sum_{i}\cfrac{Y_{\tau i}\, Y_{\mu i}^{*}}{(m_{0}/\text{TeV})^{2}}\Big |^{2}< 25\,,
\end{equation}
and thus by substituting this bound in eq.~\eqref{eq:idm} we find:
\begin{equation}
\mathrm{BR}(h \rightarrow \mu \tau)_{\rm IDM} \lesssim 10^{-7} \,  \lambda_{3}^2 \ .
\end{equation}
$\lambda_{3}$ is constrained by $h\rightarrow \gamma \gamma$,
and also by dark matter phenomenology if the dark matter particle is the neutral component of the doublet. In any case, this branching ratio is beyond any future experimental sensitivity.

\section{Conclusions} \label{sec:conc}

We have studied Higgs lepton flavour violation in the light of recent CMS results on $h\rightarrow\tau \mu$. From the effective field theory of point, we have discussed the effective operators that can give rise to HLFV, assuming that the only light degrees of freedom are those of the SM. Then, we have analyzed the different ways of obtaining these operators from a complete renormalizable theory at tree level and at one loop.

At tree level, we have listed all the topologies, containing at most two heavy new multiplets, that generate the HLFV operators. In the models obtained we have estimated the $h\rightarrow\mu\tau$ and $\tau\rightarrow \mu\gamma$ rates and have seen that in most of them, these two processes are tightly related. In the case of models containing vector-like fermions one can show that they always generate \emph{derivative} operators (what we call topologies E) which lead to non-universal and lepton flavor changing Z couplings. We showed that models containing only one vector-like multiplet (topologies E) cannot give large HLFV.

On the other hand, models containing scalar triplets, which obtain a vacuum expectation value, are constrained by the $\rho$ parameter. Models containing two vector-like leptons (topologies D) or one vector-like lepton and a scalar (topologies C) can have enhanced HLFV but then, $\tau\rightarrow \mu\gamma$ is also enhanced and current bounds forbid large HLFV. Finally, models containing, at least, one new scalar doublet (topology A, and also a slight variation containing an extra scalar, topology B) are able to yield a sizable contribution, unless the new scalar in topology B is a triplet.

When HLFV is generated at one loop, in general, it is very small, $<10^{-4}$ and typically $<10^{-7}$, as the rates are suppressed by a loop factor and constrained by CFLV, especially $\tau\rightarrow \mu\gamma$. 

We have reviewed the most popular neutrino mass models as a possible explanation of the HLFV anomaly. In most of them (Dirac masses, see-saws type-I, II and III, inverse see-saw, Zee-Babu and the Scotogenic model), HLFV appears only at one loop and suppressed by the tau mass, and, therefore, it is too small. In the case of inverse see-saw type III the fermion triplets are vector-like, but they generate HLFV at tree level via only the \emph{derivative} operator, so it can not be large. However, there are two very well-motivated models, the (general) Zee model and left-right models (with an extended scalar sector), which include an extra doublet and can explain naturally neutrino masses and in principle give a large enough $h\rightarrow \mu\tau$ rate. 

At present, neutrino oscillations are the only evidence of non-conservation of flavour in the lepton sector. 
Future data from the LHC, with a confirmation both by CMS and ATLAS of the $h\rightarrow \mu \tau$ excess would be the first indication of LFV in processes not involving neutrinos. Hopefully, this work, in which we have studied general HLFV and its connection with neutrinos, will help to pin-down the preferred models.

\section*{Acknowledgements} 
We thank Martin Hirsch and Andreas Crivellin for discussions. This work has been partially supported by the  European Union Horizon 2020 research and innovation programme under the Marie Sklodowska-Curie grant agreements No 674896 and No 690575, by the Spanish ``Ministerio de Econom\'ia y Competitividad'' under grants  FPA2014-54459-P and  FPA2014-57816-P, by the ``Generalitat Valenciana'' grants GVPROMETEOII 2014-050 and GVPROMETEOII 2014-087, and by the ``Centro de Excelencia Severo Ochoa'' Programme under grant SEV-2014-0398. All Feynman diagrams have been drawn using JaxoDraw \cite{Binosi:2003yf,Binosi:2008ig}. We also acknowledge the Higgs Tasting Workshop 2016, held at the ``Centro de Ciencias de Benasque Pedro Pascual'', Benasque (Spain), in May 15-21, 2016, where this work was finalized and for partial support for JHG. The workshop was partly supported by the European Commission under project FP7-PEOPLE-2013-CIG-631962, and by the Spanish ``Ministerio de Econom\'ia y Competitividad'' under the programme ``Centros de Excelencia Severo Ochoa'' SEV-2012-0234.

\appendix

\section{Non-diagonal Z couplings} \label{sec:nondiag}

To illustrate how non-diagonal $Z$ couplings arise from the \emph{derivative} operators in table \ref{tab:topologiesE} let us consider the topology $E_1$ in the first row. After SSB we will have a modification of the kinetic terms of the right-handed leptons $e_{R}$
\begin{equation}
\bar{e}_{R}i\slashed De_{R}+
\frac{\kappa}{v^{2}}(\bar{e}_{R}v)i\slashed D(e_{R}v)\, ,
\end{equation}
where the first term is the SM kinetic term and the second one comes from the effective operator. $\kappa$ is a Hermitian matrix in flavour, basically, the coupling of the effective operator. Typically (see appendix \ref{sec:example-derivative}) 
\begin{equation}
\kappa \sim Y_E \frac{v^2}{M_E^2} Y_E^\dagger \, ,
\end{equation}
and it is directly related to the $y_{\tau\mu}$ parametrizing HLFV (recall that for \emph{derivative} operators $y_{\tau\mu}$ is suppressed by an additional $\tau$ Yukawa coupling factor $y_\tau= m_\tau/v$ coming from the use of the equations of motion).  
\begin{equation}
y_{\tau\mu} \sim y_\tau \kappa_{\tau\mu}  \, .
\end{equation}
Notice, that the covariant derivative acts on objects with different quantum numbers. In the first term, $e_R$, is a singlet with hypercharge $-1$ while in the second term $e_R v$ is the lower component of a doublet with hypercharge $-1/2$ (the same quantum numbers of $e_L$). Then, for the first term we have
\[
\bar{e}_{R}\left(i\slashed\partial-g^{\prime}\slashed B\right)e_{R}=\bar{e}_{R}\left(i\slashed\partial-e\slashed A+\frac{es_{W}}{c_{W}}\slashed Z\right)e_{R} \, ,
\]
while for the second one we obtain
\[
\bar{e}_{R}\left(i\slashed\partial-\frac{1}{2}g^{\prime}\slashed B-\frac{1}{2}g\slashed W_{3}\right) e_R=\bar{e}_{R}\left(i\slashed\partial-e\slashed A+\left(\frac{es_{W}}{2c_{W}}-\frac{ec_{W}}{2s_{W}}\right)\slashed Z\right)e_{R}\, . 
\]
Here we used $g^{\prime}=e/c_{W}$, $g=e/s_{W}$ and 
\begin{eqnarray*}
B^\mu & = & c_{W}A^\mu-s_{W}Z^\mu\,,\\
W_{3}^\mu & = & s_{W}A^\mu+c_{W}Z^\mu \, .
\end{eqnarray*}
Adding the two terms we can write
\[
\left(1+\kappa\right)\bar{e}_{R}\left(i\slashed\partial-e\slashed A+\frac{es_{W}}{c_{W}}\slashed Z\right)e_{R}-\kappa\frac{e}{2c_{W}s_{W}}\bar{e}_{R}\slashed Ze_{R}\ .
\]
After renormalization of the first term, $e_R\rightarrow (1+\kappa)^{-1/2} e_R$ and expanding for small $\kappa$ we recover the SM coupling plus the non-diagonal $Z$ interactions
\begin{equation}
{\cal L}_{\rm ZLVF} = -\kappa\frac{e}{2c_{W}s_{W}}\bar{e}_{R}\slashed Ze_{R}\, .
\end{equation}
Notice, that the renormalization above only affects $e_{R}$ and, thus,  it does not modify charged currents at all. 

As we have seen, the lepton flavour violating neutral current interaction appears because of the mismatch between the quantum numbers of the combination of fields appearing in the new operator and the the quantum numbers of the SM  fermion fields. Using this we can generalize the procedure to all the operators in table \ref{tab:topologiesE} and obtain the results presented there, where we have also included the contributions to charged-current interactions.

\section{An example of a model generating \emph{derivative} operators}\label{sec:example-derivative}
For illustration, we will give here some details of a model giving rise to \emph{derivative} operators. We add a vector-like lepton $E=(1,-1)_{\rm F}$ (topology $E_{4a}$) of mass $M_{\rm E}$ (for simplicity we just add one vector-like lepton but the model can easily be enlarged to include several of them). The most general Yukawa Lagrangian reads
\begin{equation}
\mathcal{L}= i\overline{L}\, \slashed{D}\,L+i\overline{e_{\rm R}}\, \slashed{D}\,e_{\rm R} +\overline{E}\,(i \slashed{D}-M_{\rm E})\,E+(\overline{L} Y_e e_{\rm R}\Phi + \overline{L} Y_E E_{\rm R}\Phi +{\rm H.c.})\,,
\end{equation}
where $Y_e$ is the SM Yukawa coupling. For just one vector-like lepton, $Y_E$ is $3\times 1$ general matrix (for $n$ vector-like leptons it would be a $3\times n$ matrix and $M_E$ would be a $n\times n$ matrix).  Notice that mixed bare terms such $M \overline{e_{\rm R}}\,E_{\rm L}$ can always be reabsorbed in a redefinition of $e_{\rm R}$. We assume that $M_{\rm E}> v$, so we can integrate-out the E. Using the EOM, and substituting back in the Lagrangian we get the effective Lagrangian:
\begin{equation}
\mathcal{L_{\rm EFT}}= -(\overline{L}\,\Phi)\,Y_E \frac{1}{i \slashed{D}-M_{\rm E}}Y_E^\dagger (\Phi^\dagger \,L)\,.\label{EFTVL}
\end{equation}
Expanding the propagator of eq. \eqref{EFTVL} up to dimension 6 we find:
\begin{equation}
\mathcal{L}^{D\leq 6}_{\rm EFT}= 
(\overline{L}\,\Phi)\,\frac{Y_E Y_E^\dagger}{M_{\rm E}^2} i \slashed{D}\,(\Phi^\dagger \,L)= 
i\, \left((\overline{L}\,\Phi)\,C_E \gamma^\mu D_\mu\,(\Phi^\dagger \,L)
-D_\mu\,(\overline{L}\,\Phi)\,C_E \gamma^\mu\,(\Phi^\dagger \,L)\right)\,,
\end{equation}
where the even powers of $\slashed{D}$ vanish due to the chirality of the (Hermitian) operator.
We have also used an integration by parts to rewrite the Lagrangian in an explicitly Hermitian form and defined the $3\times 3$ matrix  matrix $C_E/\Lambda^2 =1/2\, Y_E M_E^{-2} Y_E^\dagger$ (for $n$ $E$'s $C_E$ would be given by the same expression but with $M_E$ a $n\times n$ matrix and $\Lambda$ is the lightest of the $M_E$ eigenvalues). The factor of 2 comes from the fact that we have defined the operator plus its Hermitian in eq.~\eqref{eq:hlfv-lagrangian}. Now we expand the covariant derivatives of the product of fields and use ${\rm SU(2)}$ identities like $\Phi \Phi^\dagger = \frac{1}{2}(\Phi^\dagger \Phi)+\frac{1}{2}\vec{\sigma}(\Phi^\dagger\vec{\sigma} \Phi)$ to write
\begin{eqnarray}
\mathcal{L_{\rm EFT}}^{D\leq 6}=&
\displaystyle{\frac{i}{2\Lambda^2}}
\Big[ (\overline{L}\,C_E\,\overset{\leftrightarrow }{\slashed{D}}\,L)\, \,(\Phi^\dagger\,\Phi)+( \overline{L}\,C_E\,\vec{\sigma}\,\overset{\leftrightarrow }{\slashed{D}}\,\,L)\, \,(\Phi^\dagger\,\vec{\sigma}\,\Phi)\nonumber \\
&-\,(\overline{L}\,C_E\,\gamma^\mu  \,L)\, (\Phi^\dagger \overset{\leftrightarrow }{D_\mu} \Phi)- \,(\overline{L}\,C_E\,\gamma^\mu \,\vec{\sigma} \,L)\, (\Phi^\dagger \,\vec{\sigma}\,\overset{\leftrightarrow }{D_\mu} \Phi) \Big], 
\label{EFTVLexp}
\end{eqnarray}
where $\Phi^\dagger \,\overset{\leftrightarrow }{D}_\mu \Phi \equiv \Phi^\dagger (D_\mu \Phi)- (D_\mu \Phi)^\dagger \Phi$ and $\Phi^\dagger \,\vec{\sigma}\overset{\leftrightarrow }{D}_\mu\Phi \equiv \Phi^\dagger \vec{\sigma} (D_\mu \Phi)- (D_\mu \Phi)^\dagger \vec{\sigma} \Phi$. 
The second line does not give rise to HLFV, as can be clearly seen by going to the the unitary gauge, but it gives ZLVF involving only charged leptons. A similar procedure can be used for the rest of topologies involving vector-like leptons. However, it is easier to use the mismatch of quantum numbers between the renormalizable and the non-renormalizable contributions to the kinetic terms, which immediately yield the new interactions, as sketched in appendix~\ref{sec:nondiag}.

\section{Higgs interactions in models with vector-like leptons} \label{sec:HiggsVL}

In models with vector-like leptons (singlets and doublets, a similar analysis can be done for vector-like triplets), Z-boson interactions with charged leptons can be written with complete generality in terms of mass eigenstates $E_{L\,(R)}$ as~\cite{delAguila:1998tp}:
\begin{equation}
\mathcal{L}_{Z}=\frac{g}{2c_{W}}\left(\bar{E}_{L}\gamma^{\mu}X_{L}E_{L}+\bar{E}_{R}\gamma^{\mu}X_{R}E_{R}+2s_{W}^{2}J_{EM}^{\mu}\right)Z^{\mu}\,,
\end{equation}
where we transformed from the weak basis to the mass basis, $E^{\rm weak}_{L\,(R)}= V_{L(R)}\,E_{L\,(R)}$. Notice that we have suppressed the flavour indices, $a$, in $E_{L,R}$, which run over all charged leptons, standard and heavy ones, and we have defined
\begin{equation}
\left(X_{L}\right)_{ba}=\left(V_{L}^{\dagger}\right)_{b\ell}\left(V_{L}\right)_{\text{\ensuremath{\ell a}}},\,\left(X_{R}\right)_{ba}=\left(V_{R}^{\dagger}\right)_{b\ell}\left(V_{R}\right)_{\text{\ensuremath{\ell a}}}\,,
\end{equation}
where $\ell$ run only on lepton doublets. 

The Higgs interactions read:
\begin{equation}
-\mathcal{L}_{h}\rightarrow\bar{E}_{L}V_{L}^{\dagger}Y_{E}V_{R}E_{R}h+\mathrm{H.c.}\,.
\end{equation}
It is easy to show that the Yukawa coupling can be written in the following way:
\begin{equation} \label{genH}
vy=vV_{L}^{\dagger}Y_{E}V_{R}=X_{L}D_{E}+D_{E}X_{R}-2X_{L}D_{E}X_{R}\,.
\end{equation}
This general expression simplifies in some cases. For instance if we only add vector-like singlets there will
be no right-handed doublets and therefore $X_R = 0$ and the coupling is $vy =X_L D_E$, while if
we add only vector-like doublets, then all left leptons are doublets, $X_L = I$ and
the coupling is $y=D_E- D_E X_R$~\cite{Dorsner:2015mja}.

Specifying for $y_{\mu\tau}$, we get:
\begin{align}
vy_{\mu\tau}&=(X_{L})_{\mu\text{\ensuremath{\tau}}}m_{\tau}+m_{\mu}(X_{R})_{\mu\tau}-2(X_{L}D_{E}X_{R})_{\mu\tau}\\
&\approx \frac{Y_{\mu F_1}v}{m_{F_1}}\frac{(Y)^\dagger_{F_1\tau}v }{m_{F_1}} m_{\tau}+\frac{Y_{\mu F_2}v}{m_{F_2}}\frac{(Y)^\dagger_{F_2\tau}v }{m_{F_2}} m_{\mu}- 2\frac{Y_{\text{\ensuremath{\mu}}F_1}v}{m_{F_1}}Y_{12}v\frac{(Y)^\dagger_{F_2\tau}v}{m_{F_2}}\,.
\end{align}
Similarly,
\begin{equation}
vy_{\tau\mu}= \frac{Y_{\tau F_1}v}{m_{F_1}}\frac{(Y)^\dagger_{F_1\mu}v }{m_{F_1}} m_{\mu}+\frac{Y_{\tau F_2}v}{m_{F_2}}\frac{(Y)^\dagger_{F_2\mu}v }{m_{F_2}} m_{\tau}-2\frac{Y_{\text{\ensuremath{\tau}}F_1}v}{m_{F_1}}Y_{12}v\frac{(Y)^\dagger_{F_2\mu}v}{m_{F_2}}\,.
\end{equation}
The first (second) terms in $y_{\mu\tau},\,y_{\tau\mu}$ come from \emph{derivative} operators obtained by exchange of a vector-like singlet (doublet), while the last term comes from topology D of the \emph{Yukawa} operator, where both singlets and doublets are exchanged. This term does not involve charged lepton masses and thus will typically dominate unless $Y_{12}v<m_{\tau}$.  One can also check that the contribution to $\tau\rightarrow\mu\gamma$ is always proportional to $y_{\mu\tau},\,y_{\tau\mu}$ and, therefore, will provide \emph{robust} limits on HLFV, as discussed in secs.~\ref{sec:Yukawa} and~\ref{sec:derivative}.

\bibliographystyle{JHEP}

\bibliography{HLFV}

\end{document}